\documentstyle[sprocl,epsf,amssymb]{article} 
 
\def\sst{\scriptscriptstyle}

\def\be{\begin{equation}} 
 
\def\ee{\end{equation}} 
 
\def\bea{\begin{eqnarray}} 
 
\def\eea{\end{eqnarray}}

\begin{document} 

\title 
  {THE STRUCTURE OF THE 
    NUCLEON\footnote{Supported by the Deutsche Forschungsgemeinschaft 
      (SFB 443)}}

\author 
{D. DRECHSEL} 
\address 
{ Institut f\"ur Kernphysik, Universit\"at Mainz \\ 
55099 Mainz, Germany \\ Email: drechsel@kph.uni-mainz.de \\ 
Homepage: http://www.kph.uni-mainz.de/T/}

\maketitle\abstracts{These lectures give an introduction to the 
  structure of the nucleon as seen with the electromagnetic probe. 
  Particular emphasis is put on the form factors, the strangeness 
  content, Compton scattering and polarizabilities, pion photo- and 
  electroproduction, the spin structure and sum rules. The existing 
  data are compared to predictions obtained from chiral perturbation 
  theory, dispersion theory and effective Lagrangians.}

 
\section{Introduction} 
Nucleons are composite systems with many internal degrees of freedom. 
The constituents are quarks and gluons, which are bound by 
increasingly strong forces if the momentum transfer decreases towards 
the GeV region. The ``running'' coupling constant of the strong 
interaction, $\alpha_s(Q^2)$ in fact diverges if $Q^2$ approaches 
$\Lambda^2_{QCD}\approx(200~$MeV/c)$^2$ corresponding to a scale in 
space of about 1~fm. This is the realm of nonperturbative quantum 
chromodynamics (QCD), where confinement plays a major role, and quarks 
and gluons cluster in color neutral objects. Such correlations between 
the constituents have the consequence that nucleons in their natural 
habitat, i.e. at the confinement scale, have to be described by 
hadronic degrees of freedom rather than quarks and gluons. 
 
QCD is a nonlinear gauge theory developed on the basis of massless
quarks and gluons~\cite{QCD73}. The interaction among the gluons gives
rise to the nonlinearity, and the interaction among the quarks is
mediated by the exchange of gluons whose chromodynamic vector
potential couples to the vector current of the quarks. If massless
particles interact by their vector current, their helicity remains
unchanged. In practice one has to restrict this discussion to $u$, $d$
and $s$ quarks with masses $m_u\approx5$~MeV, $m_d\approx9$~MeV and
$m_s\approx175$~MeV, which are all small at the mass scale of the
nucleon. These quarks can be described by SU(3)$_R\otimes$ SU(3)$_L$
as long as right and left handed particles do not interact, which is
what happens if the helicity is conserved. By combining right and left
handed currents, one obtains the vector currents $J_{\mu}^a$ and the
axial vector currents $J_{5\mu}^a$,
\begin{equation} 
J_{\mu}^a=\bar{q}\gamma_{\mu}\frac{\lambda^a}{2}q\ , 
\ \ \ J_{5\mu}^a=\bar{q}\gamma_{\mu}\gamma_5\frac{\lambda^a}{2}q\ , 
\label{intro0} 
\end{equation} 
where $q$ are Dirac spinors of the massless and point-like light 
quarks and $\gamma_{\mu},\gamma_5$ the appropriate Dirac matrices. The 
quantities $\lambda^a$, $a=1\ ...\ 8$ denote the Gell-Mann matrices of 
SU(3) describing the flavor structure of the 3 light quarks. It is 
often convenient to introduce the unit matrix $\lambda^0$ in addition 
to these matrices. 
 
In the context of these lectures we shall only need the ``neutral'' 
currents corresponding to $\lambda=3$, 8 and 0, which have a diagonal 
form in the standard representation. The photon couples to quarks by 
the electromagnetic vector current $J_{\mu}^{em}\sim 
J_{\mu}^{(3)}+\frac{1}{\sqrt{3}}J_{\mu}^{(8)}$, corresponding to 
isovector and isoscalar interactions respectively. The weak neutral 
current mediated by the $Z^0$ boson couples to the 3rd, 8th and 
0th components of both vector and axial currents. While the 
electromagnetic current is always conserved, 
$\partial^{\mu}J_{\mu}^{em}=0$, the axial current is only conserved in 
the limit of massless quarks. In this limit there exist conserved 
charges $Q^a$ and axial charges $Q_5^a$, which are connected by 
current algebra, 
\begin{equation} 
[Q^a,Q^b] = i f^{abc} Q^c \ , 
\ \ [Q^a_5,Q^b_5] =  i f^{abc} Q^c \ , 
\ \ [Q^a_5,Q^b] = i f^{abc} Q_5^c\ , 
\end{equation} 
with $f^{abc}$ the structure constants of SU(3). Such relations were 
an important basis of low energy theorems (LET), which govern the low 
energy behavior of (nearly) massless particles. 
 
The puzzle we encounter is the following: The massless quarks 
appearing in the QCD Lagrangian conserve the axial currents but 
the nucleons as their physical realizations are massive and 
therefore do not conserve the axial currents. The puzzle was 
solved by Goldstone's theorem. At the same time as the ``3 quark 
system'' nucleon becomes massive by means of the QCD interaction, 
the vacuum develops a nontrivial structure due to finite 
expectation values of quark-antiquark pairs (condensates 
$\langle\bar{q}q\rangle$), and so-called Goldstone bosons are 
created, $\bar{q}q$ pairs with the quantum numbers of pseudoscalar 
mesons.  These Goldstone bosons are massless, and together with 
the massive nucleons they act such that chirality is locally 
conserved.  This mechanism can be compared to the local gauge 
symmetry of quantum electrodynamics, which is based on the fact 
that both (massless) photon and (massive) matter fields have to be 
gauge transformed. 
 
\begin{figure}[h] 
\label{FIGintro1} 
\centerline{\epsfysize=8cm \epsfbox{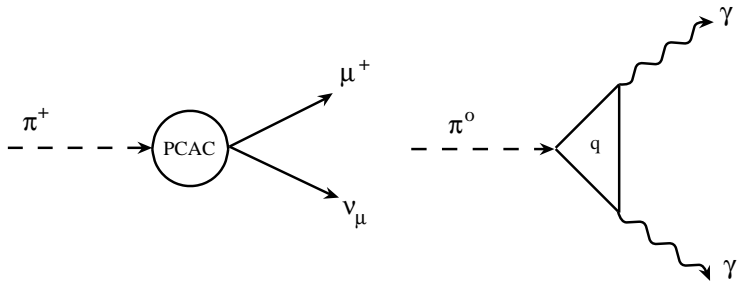}} 
\vspace{-4.0cm} 
\caption 
{The 4-divergence of the axial current (PCAC) responsible for charged 
  pion decay, and the axial anomaly visualized by an intermediate 
  quark triangle describing neutral pion decay.} 
\end{figure}

In QCD the chiral symmetry is definitely broken by the small but 
finite quark masses. As a consequence also the physical ``Goldstone 
bosons'', in particular the pions, acquire a finite mass $m_{\pi}$, 
which is generally assumed (though not proven) to follow the 
Gell-Mann-Oakes-Renner relation 
\begin{equation} 
m_{\pi}^2f_{\pi} = -(m_u+m_d)\langle\bar{q}q\rangle\ , 
\label{intro2} 
\end{equation} 
with the condensate $\langle\bar{q}q\rangle\approx 
-(225~{\mbox{MeV}})^3$, and $f_{\pi}\approx93~$MeV the pion decay 
constant. Since the pions are now massive, the corresponding axial 
currents are no longer conserved and the 4-divergence of the axial 
current becomes 
\begin{equation} 
\partial^{\mu}J_{5\mu}^a\approx -f_{\pi}m_{\pi}^2\phi_{\pi}^a\ , 
\label{intro3} 
\end{equation} 
where $\phi_{\pi}^a$ describes the local field of charged pions $(a=1$ 
 and 2). In other words the weak decays 
\begin{equation} 
  \pi^+\rightarrow\mu^++\nu_{\mu}\ \ \ {\mbox{and}}\ \ \ 
  \pi^-\rightarrow \mu^-+\bar{\nu}_{\mu} 
\label{intro4} 
\end{equation} 
proceed via coupling to the axial current 
(Fig.~\ref{FIGintro1}).The pion and its axial current disppear 
from the hadronic world and leave the (hadronic) vacuum behind. In 
particular we note that a finite value of the divergence of 
Eq.~(\ref{intro3}) has 3 requirements: the decay of the pion can 
take place, the pion mass is finite, and a local pion field 
exists. 
 
While the charged pions decay weakly with a life-time of 
$2.6\cdot10^{-8}$~sec, the neutral pion decays much faster, in 
$8.4\cdot10^{-17}$~sec, by means of the electromagnetic interaction, 
\begin{equation} 
  \pi^0\rightarrow\gamma+\gamma\ . 
\end{equation} 
Again axial current disappears, corresponding to 
\begin{equation} 
  \partial^{\mu}J_{5\mu}^3 = \frac{\alpha_{fs}}{\pi}\ 
  \vec{E}\cdot\vec{B}\ , 
\label{intro6} 
\end{equation} 
where $\alpha_{fs} = e^2/4\pi$ is the fine structure constant, and 
$\vec{E}$ and $\vec{B}$ are the electromagnetic fields. We note that 
two electromagnetic fields have to participate, because two photons 
are created, and that they have to be combined as a pseudoscalar, 
because the pseudoscalar pion disappears. The transition of 
Eq.~(\ref{intro6}) can be visualized by the intermediate quark 
triangle of Fig.~\ref{FIGintro1}. It is called the ``triangle 
anomaly'', because such transitions cannot exist in classical theories 
but only occur in quantum field theories via the renormalization 
procedure. Such terms are also predicted on general grounds 
(Wess-Zumino-Witten term). We note in passing that a similar anomaly 
is obtained in QCD by replacing the electromagnetic fields by the 
corresponding color fields, $\vec{E}_c$ and $\vec{B}_c$, $\alpha_{fs}$ 
by the strong coupling $\alpha_s$, and by an additional factor 3 for 
u, d, and s quarks, 
\begin{equation} 
  \partial^{\mu}J_{5\mu}^{0} = 3\ \frac{\alpha_s}{\pi}\ 
  \vec{E}_c\cdot\vec{B}_c \ . 
\label{intro7} 
\end{equation} 
As a consequence the component $J_{5\mu}^{0}$ is not conserved, 
not even in the case of massless quarks (``$U_A(1)$ anomaly''). 
 
Unfortunately, no ab-initio calculation can yet describe the 
interesting but complicated world of the confinement region. In 
principle, lattice gauge theory should have the potential to describe 
QCD directly from the underlying Lagrangian. However, these 
calculations have yet to be restricted to the ``quenched 
approximation'', i.e. initial configurations of 3 valence quarks. This 
is a bad approximation for light quarks, because the Goldstone 
mechanism creates plenty of sea quarks, and therefore the calculations 
are typically performed for massive quarks, $m_q\approx100$~MeV, and 
then extrapolated to the small $u$ and $d$ quark masses. In this way 
one obtains reasonable values for mass ratios of hadrons and 
qualitative predictions for electromagnetic properties.  However, some 
doubt may be in order whether such procedure will describe the typical 
threshold behavior of pionic reactions originating from the Goldstone 
mechanism. 
 
A further ``ab initio'' calculation is chiral perturbation theory
(ChPT), which has been established by Weinberg in the framework of
effective Lagrangians and put into a systematic perturbation theory by
Gasser and Leut-wyler~\cite{Gas84}. Based on the Goldstone mechanism,
the threshold interaction of pions is weak both among the pions and
with nucleons, and furthermore the pion mass is small and related to
the small quark masses $m_u$ and $m_d$ by Eq.~(\ref{intro2}). As a
consequence ChPT is a perturbation in a parameter
$p:=(p_1,p_2,...;m_u,m_d)$, where $p_i$ are the external 4-momenta of
a particular (Feynman) diagram. (Note that also the time-like
component, the energy, is small at threshold because of the small
mass!). This theory has been applied to photoinduced reactions by
Bernard, Kaiser, Mei{\ss}ner and others~\cite{Ber95} over the past decade.
As a result several puzzles have been solved and considerable insight
has been gained. There exists, however, the problem that ChPT cannot
be renormalized in the ``classical'' way by adjusting a few parameters
to the observables. Instead the renormalization has to be performed
order by order, the appearing infinities being removed by counter
terms.  This procedure gives rise to a growing number of low energy
constants (LECs) describing the strength of all possible effective
Lagrangians consistent with QCD, at any given order of the
perturbation series.  These LECs, however, cannot (yet) be derived
from QCD but have to be fitted to the data, which leads to a
considerable loss of predictive powers in the higher orders of
perturbation. A further problem arises in the nucleonic sector due to
the nucleon's mass $M$, which is of course not a small expansion
parameter. The latter problem has been overcome by heavy baryon ChPT
(HBChPT), a kind of Foldy-Wouthuysen expansion in $M^{-1}$. The
solution is achieved, however, at the expense of going from an
explicitly relativistic field theory to a nonrelativistic scheme.
 
Beside lattice gauge theory and ChPT, which are in principle directly 
based on QCD, there exists a host of QCD inspired models, which we 
shall not discuss at this point but occasionally refer to at later 
stages.

\section{KINEMATICS} 
\begin{figure}[t] 
\label{FIGkin2} 
\centerline{\epsfxsize=3cm \epsfbox{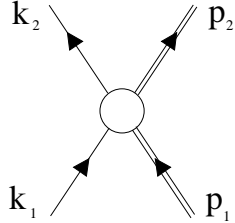}} 
\caption 
{The reaction $k_1+p_1\rightarrow k_2+p_2$. The 4-momenta $p_1$ and 
  $p_2$ describe a nucleon in the initial and final states 
  respectively, while $k_1$ and $k_2$ stand for a lepton.} 
\end{figure}

Let us consider the kinematics of the reaction 
\begin{equation} 
e(k_1)+N(p_1)\rightarrow e(k_2)+N(p_2)\ , 
\label{KIN1} 
\end{equation} 
with $k_1=(\omega_1,\vec{k}_1)$ and $p_1=(E_1,\vec{p}_1)$ denoting the 
four-momenta of an electron $e$ and a nucleon $N$ in the initial 
state, and corresponding definitions for the final state 
(Fig.~\ref{FIGkin2}). These momenta fulfil the on-shell conditions 
\begin{equation} 
p_1^2=p_2^2=M^2\ ,\ \ \ k_1^2=k_2^2=m^2\ , 
\label{KIN2} 
\end{equation} 
and furthermore conserve total energy and momentum, 
\begin{equation} 
k_1+p_1=k_2+p_2\ . 
\label{KIN3} 
\end{equation} 
%
 
If we also assume parity conservation, the scattering amplitudes 
should be Lorentz invariants depending on the Lorentz scalars that can 
be constructed from the four-momenta. By use of Eqs.~(\ref{KIN2}) 
and~(\ref{KIN3}) it can be shown that there exist only two independent 
Lorentz scalars, corresponding to the fact that the kinematics of 
Eq.~(\ref{KIN1}) is completely described by, e.g., the lab energy of 
the incident electron, $\omega_L$, and the scattering angle 
$\Theta_L$. In order to embed relativity explicitly, it is useful to 
express the amplitudes in terms of the 3 Mandelstam variables 
\begin{equation} 
s=(k_1+p_1)^2\ \ ,\ \ t=(k_2-k_1)^2\ \ ,\ \ u=(p_2-k_1)^2\ . 
\end{equation} 
Since only two independent Lorentz scalars exist, these variables have 
to fulfil an auxiliary condition, which is 
\begin{equation} 
s+t+u=2\ (m^2+M^2)\ . 
\label{KIN5} 
\end{equation} 
In the $cm$ frame, the 3-momenta of the particles cancel and 
$s=(\omega_{cm}+E_{cm})^2=W_s^2=W^2$, where $W$ is the total energy in 
that frame. Furthermore, the initial and final energies of each 
particle are equal, hence $t=-(\vec{k}_2-\vec{k}_1)^2_{cm}=-\vec{q}\ 
^2_{cm}$, where $\vec{q}_{cm}$ is the 3-momentum transfer in the $cm$ 
system.  From these definitions it follows that $s\ge (m+M)^2$ and 
$t\le 0$ in the physical region. Since $s$ is Lorentz invariant, the 
threshold energy $\omega_{lab}$ can be obtained by comparing $s$ as 
expressed in the $lab$ and $cm$ frames. Moreover, in a general frame 
$t=(k_2-k_1)^2=q^2<0$ describes the square of 4-momentum of the 
virtual photon $\gamma^{\ast}$, exchanged in the scattering process 
(``space-like photon'').  Since $t$ is negative in the physical region 
of electron scattering, we shall define the positive number $Q^2=-q^2$ 
for further use. We also note that in pair annihilation, 
$e^+e^-\rightarrow\gamma^{\ast}$, the square of 4-momentum is 
positive, $q^2=m_{\gamma^{\ast}}^2>0$ (``time-like photon''). 
 
The above equations can be easily applied to Compton scattering, 
\begin{equation} 
\gamma(k_1)+N(p_1)\rightarrow\gamma(k_2)+N(p_2)\ , 
\end{equation} 
by replacing $m$ by zero, the mass of a real photon, and to 
virtual Compton scattering (VCS), 
\begin{equation} 
\gamma{\ast}(k_1)+N(p_1)\rightarrow\gamma(k_2)+N(p_2)\ , 
\end{equation} 
by replacing $m^2\rightarrow k_1^2=q^2<0$. Due to the spins of 
photon and nucleon, several Lorentz structures appear in the 
scattering amplitude, and each of these structures has to be 
multiplied by a scalar function depending in the most general case 
on 3 variables, $F=F(s,t,Q^2)$. 
 
Another generalization occurs if the nucleon is excited in the 
scattering process, in which case $p_2^2=(M^{\ast})^2>M^2$ becomes an 
additional variable. Introducing the Bjorken variable $x=Q^2/2p_1\cdot 
q$ we find that $x=1$ corresponds to elastic scattering, while 
inelastic scattering is described by values $0\le x<1$. 
 
\begin{figure}[htbp] 
\centerline{\epsfxsize=10cm \epsfbox{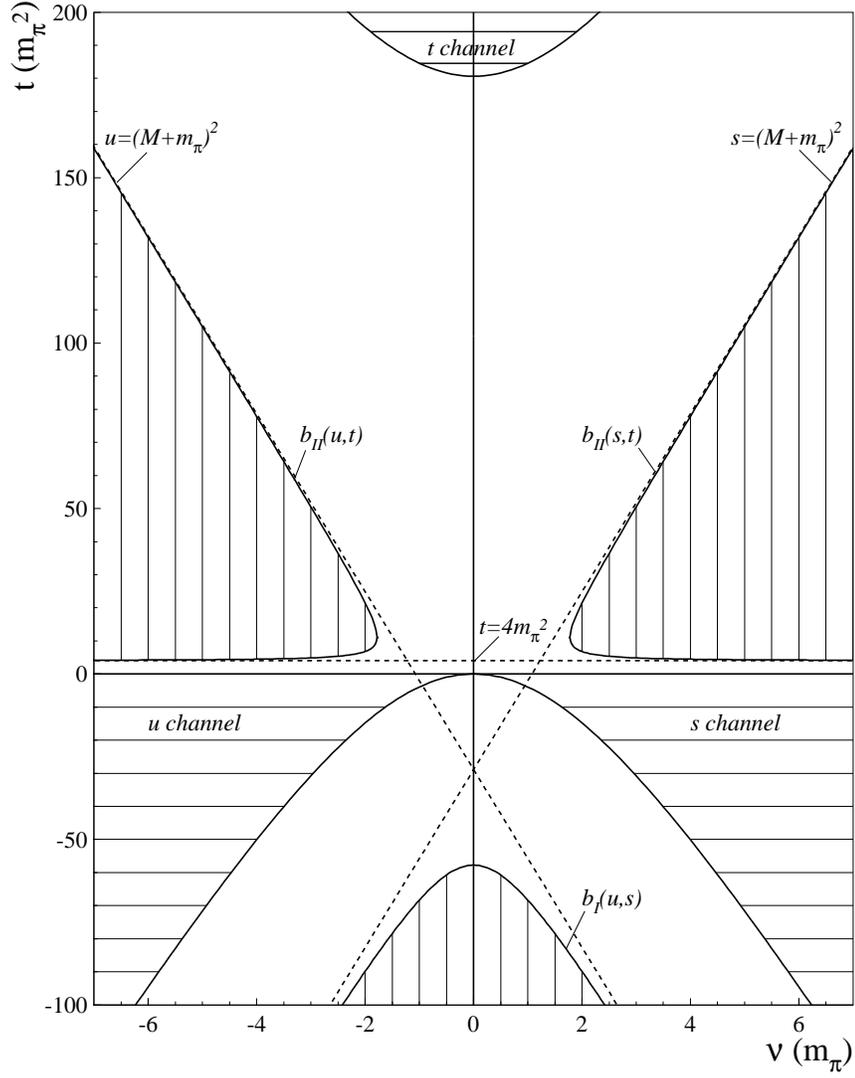}} 
\caption 
{The Mandelstam plane for Compton scattering, with the crossing 
  symmetrical variable $\nu=(s-u)/4M$ and $t$ as orthogonal 
  coordinates. The horizontally hatched areas are the physically 
  allowed regions for s, t, and u channel kinematics. The scattering 
  amplitudes become complex if particle production is allowed, i.e. 
  for $t\ge 4m_{\pi}^2$, and $s$ or $u\ge (M+m_{\pi})^2$. As a 
  consequence the scattering amplitudes are real inside the triangle 
  formed by the dashed lines near the origin.} 
\label{KIN.fig2} 
\end{figure} 
 
For further use we shall acquaint ourselves with the Mandelstam 
plane for (real) Compton scattering, as shown by 
Fig.~\ref{KIN.fig2}. Due to the symmetry of the Mandelstam 
variables, the figure can be constructed on the basis of a 
triangle with equal sides and heights equal to $2M^2$ according to 
Eq.~(\ref{KIN5}) for $m=m_{\gamma}=0$. The axes $s=0$, $t=0$, and 
$u=0$ are then obtained by drawing straight lines through the 
sides of the triangle. The physically allowed region for 
$k_1+p_1\rightarrow k_2+p_2$ is given by the horizontally hatched 
area called ``s channel'' with $s\ge M^2$ and $t\le 0$. If we 
replace $p_1\rightarrow -p_1$ and $p_2\rightarrow -p_2$ in 
Eq.~(\ref{KIN3}), we obtain the ``u channel'' reaction 
$k_1+p_2\rightarrow k_2+p_1$ given by the horizontally hatched 
area to the left.

\begin{figure}[htbp] 
\centerline
{\epsfxsize=11cm \epsfbox{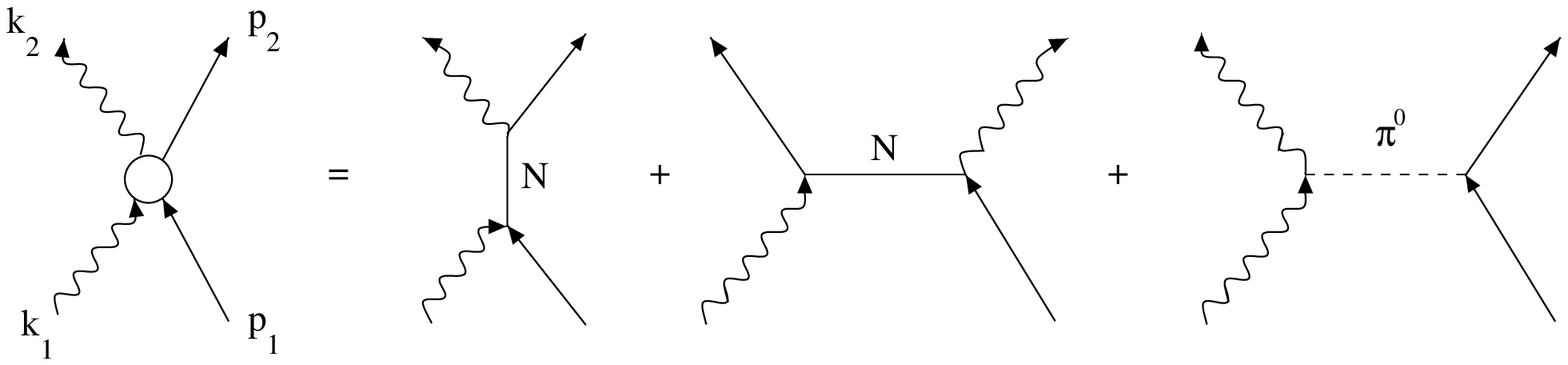}} 
\caption 
{The pole terms contributing to Compton scattering. From left to right
  on the $rhs$ the direct nucleon pole term, the crossed nucleon pole
  term, and the $\pi^0$ pole term.}
\label{FIGkin4} 
\end{figure} 

Finally, if we look at Fig.~\ref{FIGkin2} from the left side, we 
obtain the $t$ channel reaction 
$\gamma(k_1)+\gamma(-k_2)\rightarrow N(p_1)+\bar{N}(-p_2)$, which 
corresponds to the replacements $k_2\rightarrow -k_2$ and 
$p_1\rightarrow -p_1$ and is physically observable for $t>2M^2$ 
(hatched area at top of Fig.~\ref{KIN.fig2}). Referring again to 
the $s$ channel, the boundaries of the physical region correspond 
to the scattering angles 0 and 180$^{\circ}$. The former case 
leads to zero momentum transfer, i.e. the line $t=0$, the latter 
case to the hyperbolic boundary of the region at negative $t$ 
values. The u-channel region is then simply obtained by a 
reflection of the figure at the line $s=u$ given by the $t$ axis. 
Finally, the boundary of the t-channel region is given by the 
upper branch of the hyperbola, separated from the lower one by 
$4M^2$. 
 
Still in the context of Compton scattering, Fig.~\ref{FIGkin4} 
shows the Born diagrams (tree graphs) contributing to the 
reaction. In order of appearance on the $rhs$, we find the direct, 
the crossed, and the $\pi^0$ pole terms, exhibiting pole 
structures as $(s-M^2)^{-1}$, $(u-M^2)^{-1}$ and 
$(t-m_{\pi}^2)^{-1}$ respectively. Except for the origin at 
$s=u=M^2$ (``scattering'' of photons with zero momentum), these 
poles are situated on straight lines outside of the physical 
regions. However, photon scattering at small energies is obviously 
dominated by the poles at $s=M^2$ and $u=M^2$. The ``low-energy 
theorem'' asserts that for a particle with charge $e$ and mass 
$M$, the scattering amplitude behaves as 
$T=-\frac{e^2}{M}+O(\omega_{cm}^2)$, where $e^2/4\pi\approx 
1/137$. It is derived on the basis that (i) only the Born terms 
have pole singularities for $\omega_{cm}\rightarrow 0$, which 
results in the Thomson amplitude $(-e^2/M)$, and (II) gauge 
invariance or current conservation, which allows one to express 
the next-to-leading-order terms in $\omega_{cm}$ by the Born contributions. 
Therefore, the internal structure (polarizability) of the system 
enters only in terms of relative order $\omega_{cm}^2$, i.e. is 
largely suppressed near threshold. 
 
If the energy of the photon is sufficient to produce a pion,
$\sqrt{s}>M+m_{\pi}$, Compton scattering competes with the much
stronger hadronic reactions and becomes complex. The same is true in
the $t$ channel, whenever the two photons carry more energy than
$\sqrt{t}=2m_{\pi}$. Therefore the Compton amplitudes are only real in
an area around the origin $(s=u=M^2,\ t=0)$, i.e. in the triangle
shaped by the dashed lines in Fig.~\ref{KIN.fig2}. Due to this
reality relation, however, the Compton amplitudes can be analytically
continued into the unphysical region, and information from the
different physical regions can be combined to construct a common
amplitude for the whole Mandelstam plane.  Summarizing the role of the
singularities for the specific reaction of Compton scattering we find:
(I) The nucleon poles in the direct and crossed Born graphs, at
$s=M^2$ and $u=M^2$, which are close to and therefore important near
threshold, (II) the pion pole term at $t=m_{\pi}^2$ and a branch cut
starting at $t=4m_{\pi}^2$ due to the opening of the $2\pi$ continuum,
which affect the forward amplitude at any energy and (III) the opening
of hadronic channels at $s,u> (M+m_{\pi})^2$, which lead to a complex
amplitude and a much enhanced Compton cross section, particularly
near resonances at $s=M_{\mbox{\scriptsize{res}}}^2$.
 
Let us finally consider the spin degrees of freedom of the involved 
particles. A virtual photon with momentum $\vec{q}$ carries a 
polarization described by the vector potential $\vec{A}$, which has 
both a transverse part, $\vec{A}_T\perp \vec{q}$, as in the case of a 
real photon, and a longitudinal component $\hat{q}\cdot\vec{A}$, which 
is related to the time-like component $A_0$ by current conservation, 
$q\cdot A=q_0 A_0-\vec{q}\cdot\vec{A}=0$. As a consequence the cross 
section for the reaction of Eq.~(\ref{KIN1}) takes the (somewhat 
symbolical) form 
\begin{equation} 
  \frac{d\sigma}{d\Omega} = \Gamma (\sigma_T+\varepsilon\sigma_L)\ , 
\label{KIN8} 
\end{equation} 
where $\Gamma$ describes the flux of the virtual photon spectrum, and 
$\sigma_T$ and $\sigma_L$ the transverse and longitudinal cross 
sections respectively. The so-called transverse polarization 
$\varepsilon$ of the virtual photon field is given by kinematical 
quantities only, which can be varied such that the partial cross 
sections remain constant. In this way the two partial cross sections 
can be separated by means of a ``Rosenbluth plot''. 
 
Concerning the electron, we shall assume that it is highly 
relativistic, hence its spin degree of freedom will be described by 
the helicity $h=\vec{s}\cdot\hat{k}=\pm\frac{1}{2}$, the projection of 
the spin $\vec{s}$ on the momentum vector $\vec{k}$. As long as the 
interaction is purely electromagnetic, a polarization of the 
electron alone does not change the structure of the cross section, 
Eq.~(\ref{KIN8}). However, new structures appear if both electron and 
nucleon are polarized. In particular the reaction 
$\vec{e}+\vec{N}\rightarrow$ anything is described by the cross 
section~\cite{Dre95} 
\begin{equation} 
  \frac{d\sigma}{d\Omega} = \Gamma 
  [\sigma_T+\varepsilon\sigma_L+P_eP_x 
  \sqrt{2\varepsilon(1-\varepsilon)}\ \sigma'_{LT} 
  +P_eP_z\sqrt{1-\varepsilon^2}\ \sigma'_{TT}] \ , 
\label{KIN9} 
\end{equation} 
where $P_e=2h=\pm1$ refers to the helicity of the electron, and $P_z$ 
and $P_x$ are the longitudinal and transverse polarizations of the 
nucleon defined by the momentum of the virtual photon and an axis 
perpendicular to that direction (note: $P_x$ lies in the scattering 
plane of the electron and takes positive values on the side of the 
scattered electron). 
 
In a more general experiment with production of pseudoscalar mesons, 
e.g. pions, 
\begin{equation} 
\vec{e}+\vec{N}\rightarrow e'+N'+\pi\ , 
\end{equation} 
up to 18 structure functions can be defined~\cite{Dre92}, and this 
number increases further when higher spins are involved, e.g. if the 
electron is scattered on a deuteron target or if a vector particle 
(real photon, $\rho$ or $\omega$ meson etc.) is produced.

\section{FORM FACTORS} 
Consider the absorption of a virtual photon with four-momentum q at an 
hadronic vertex. If the hadron stays intact after this process, i.e. 
in the case of elastic lepton scattering, the photon probes the 
expectation value of the hadronic vector current. If 
moreover the hadron is a scalar or pseudoscalar particle, the 
vector current has to be proportional to the two independent 
combinations of the 3 external four-momenta. Choosing $q=p_2-p_1$ and 
$P=(p_1+p_2)/2$ as the independent vectors, 
\begin{equation} 
  J_{\mu} := \langle p_2\mid J_{\mu}\mid p_1\rangle = F_1
  \frac{P_{\mu}}{m} + F_2 \frac{q_{\mu}}{m} \ .
\label{FF1} 
\end{equation} 
In this way we define two form factors, $F_1$ and $F_2$, which have to 
be scalars and as such may be expressed by functions of the 
independent Lorentz scalars that can be constructed. It is again a 
simple exercise to show that there exists only one independent scalar, 
e.g. $Q^2=-q^2$, because $P\cdot q=0$ and $P^2=m^2-\frac{1}{4}q^2$ in 
the case of elastic scattering off a particle with mass $m$. 
 
Next we can exploit the fact that the vector current of 
Eq.~(\ref{FF1}) is conserved, which follows from gauge invariance. The 
result is 
\begin{equation} 
0 = q_{\mu} J^{\mu} = F_1\frac{p_2^2-p_1^2}{2m}+F_2\frac{q^2}{m}  \ . 
\end{equation} 
Since $p_1^2=p_2^2=m^2$ for on-shell particles, the first term is zero 
and hence $F_2$ has to vanish identically. Therefore the vector 
current of, e.g., an on-shell pion has to take the form 
\begin{equation} 
J^{\mu}(\pi) =  \frac{p_1^{\mu}+p_2^{\mu}}{2m_{\pi}}\ F_{\pi}(Q^2) \ . 
\end{equation} 
The form factor is normalized to $F_{\pi}(0)=e_{\pi}$, here and in the 
following in units of the elementary charge e. In this way we obtain, in 
the static limit $q_{\mu}\rightarrow 0$ and $p_{2\mu}\rightarrow 
p_{1\mu}\Rightarrow (m_{\pi},\vec{0})$, the result $J_{\mu}\Rightarrow 
(e_{\pi},\vec{0})$ for a charge $e_{\pi}$ at rest. 
 
The situation is more complicated in the case of a particle with a 
spin like the nucleon, because now the independent momenta $q$ and $P$ 
can be combined with the familiar 16 independent $4\times 4$ matrices 
of Dirac's theory: 1 (scalar), $\gamma_5$ (pseudoscalar), 
$\gamma_{\mu}$ (vector), $i\gamma_5\gamma_{\mu}$ (axial vector), and 
$\sigma_{\mu\nu}$ (antisymmetrical tensor). It is straightforward but 
somewhat tedious to show that the most general vector current of a 
spin-1/2 particle has to take the form 
\begin{equation} 
J_{\mu} :=  \langle p_2\mid J_{\mu}\mid p_1\rangle = \bar{u}_{p_2} \left 
    (F_1\gamma_{\mu}+i\frac{F_2}{2m} \sigma_{\mu\nu}q^{\nu} \right ) 
  u_{p_1}\ , 
\label{FF4} 
\end{equation} 
where $u_{p1}$ and $u_{p2}$ are the 4-spinors of the nucleon in the 
initial and final states respectively. The first structure on the 
$rhs$ is the Dirac current, which appears with the Dirac form factor 
$F_1$. The second term reflects the fact that due to its internal 
structure the particle acquires an anomalous magnetic moment $\kappa$, 
which appears with the Pauli form factor $F_2$. These form factors are 
normalized to $F_1^p(0)=1$, $F_2^p (0)=\kappa_p=1.79$ and $F_1^n 
(0)=0$, $F_2^n (0)=\kappa_n=-1.91$ for proton and neutron 
respectively. 
 
From the analogy with nonrelativistic physics, it is seducing to
associate the form factors with the Fourier transforms of charge and
magnetization densities. The problem is that a calculation of the
charge distribution $\rho(\vec{r})$ involves a 3-dimensional Fourier
transform of the form factor as function of $\vec{q}$, while in
general the form factors are functions of $Q^2=\vec{q}\ ^2-\omega^2$.
However, there exists a special Lorentz frame, the Breit or brickwall
frame, in which the energy of the virtual photon vanishes. This can be
realized by choosing, e.g., $\vec{p}_1=-\vec{q}\ /2$ and
$\vec{p}_2=+\vec{q}\ /2$ leading to $E_1=E_2=(m^2+\vec{q}\ 
^2/4)^{1/2}$, $\omega=0$, and $Q^2=\vec{q}\ ^2$.
 
In that frame the vector current takes the form 
\begin{equation} 
J_{\mu}  = \left (G_E(Q^2)\ ,\ 
  i\frac{\vec{\sigma}\times\vec{q}}{2m} G_M(Q^2)\right ) \ , 
\end{equation} 
where $G_E$ stands for the time-like component of $J_{\mu}$ and hence
is identified with the Fourier transform of the electric charge
distribution, while $G_M$ appears with a structure typical for a
static magnetic moment and hence is interpreted as Fourier transform
of the magnetization density. The two ``Sachs form factors'' $G_E$ and
$G_M$ are related to the Dirac form factors by~\cite{Sac62}
\begin{equation} 
G_E (Q^2) = F_1 (Q^2) - \tau F_2 (Q^2)\ ,\ \ \ 
G_M (Q^2) = F_1 (Q^2) + F_2 (Q^2)\ , 
\label{FF6} 
\end{equation} 
where $\tau=Q^2/4m^2$ is a measure of relativistic (recoil) effects. 
While Eq.~(\ref{FF6}) is taken as a general, covariant definition, the 
Sachs form factors can only be Fourier transformed in a special frame, 
namely the Breit frame, with the result 
\begin{eqnarray} 
G_E (\vec{q}\ ^2)& = &\int\rho(\vec{r}) e^{i\vec{q}\cdot\vec{r}} 
d^3\vec{r} \nonumber \\ 
& = & \int\rho(\vec{r})d^3\vec{r}-\frac{\vec{q}\ ^2}{6} 
\int\rho(\vec{r})\vec{r}\ ^2 d^3\vec{r} +\ ...\ , 
\end{eqnarray} 
where the first integral yields the total charge in units of $e$, i.e. 
1 for the proton and 0 for the neutron, and the second integral 
defines the square of the electric $rms$ radius, $\langle r^2\rangle_E 
:= r_E^2$ of the particle. The interpretation of $G_E$ in terms of the 
charge distribution has recently been discussed again~\cite{Isg98}. 
 
We note that each value of $Q^2$ requires a particular Breit frame. 
Therefore, information has to be compiled from an infinity of 
different frames, which is then used as input for the Fourier integral 
for $\rho(\vec{r})$ in terms of $G_E(\vec{q}\ ^2)$.  Therefore, the 
density $\rho(\vec{r})$ is not an observable that we can ``see'' in 
any particular Lorentz frame but only a mathematical construct in 
analogy to a ``classical'' charge distribution. The problem is, of 
course, that due to the small mass of an ``elementary'' particle, 
recoil effects (measured by $\tau$) and size effects (measured by 
$\langle r^2\rangle$) become comparable and cannot be separated in a 
unique way.  This situation is numerically quite different in the case 
of a heavy nucleus for which the size effects dominate the recoil 
effects by many orders of magnitude! 
 
The two Sachs form factors may be determined from the differential 
cross section 
\begin{equation} 
  \frac{d\sigma}{d\Omega} =\sigma_{\sst{\mbox{Mott}}} \left ( 
    \frac{G_E^2+\tau G_M^2}{1+\tau}+2\tau\tan^2\frac{\theta}{2}G_M^2 
  \right ) 
\end{equation} 
by means of a ``Rosenbluth plot'', showing the cross section as 
function of $\tan^2\frac{\theta}{2}$ for constant $Q^2$. The data 
should lie on a straight line with a slope $2\tau G_M^2$, and the 
extrapolation to $\tau=0$ will determine the electric form 
factor $G_E$. Unfortunately, the Rosenbluth plot has a limited range 
of applicability. For decreasing $Q^2$, also $\tau$ and the slope 
become small and the error bars on $G_M^2$ increase. Large $Q^2$, on 
the other hand side, lead to a small electric contribution $\sim 
G_E^2/\tau$ with large errors for the electric form factor.

\begin{figure}[htpb] 
\centerline{\epsfxsize=11cm \epsfbox{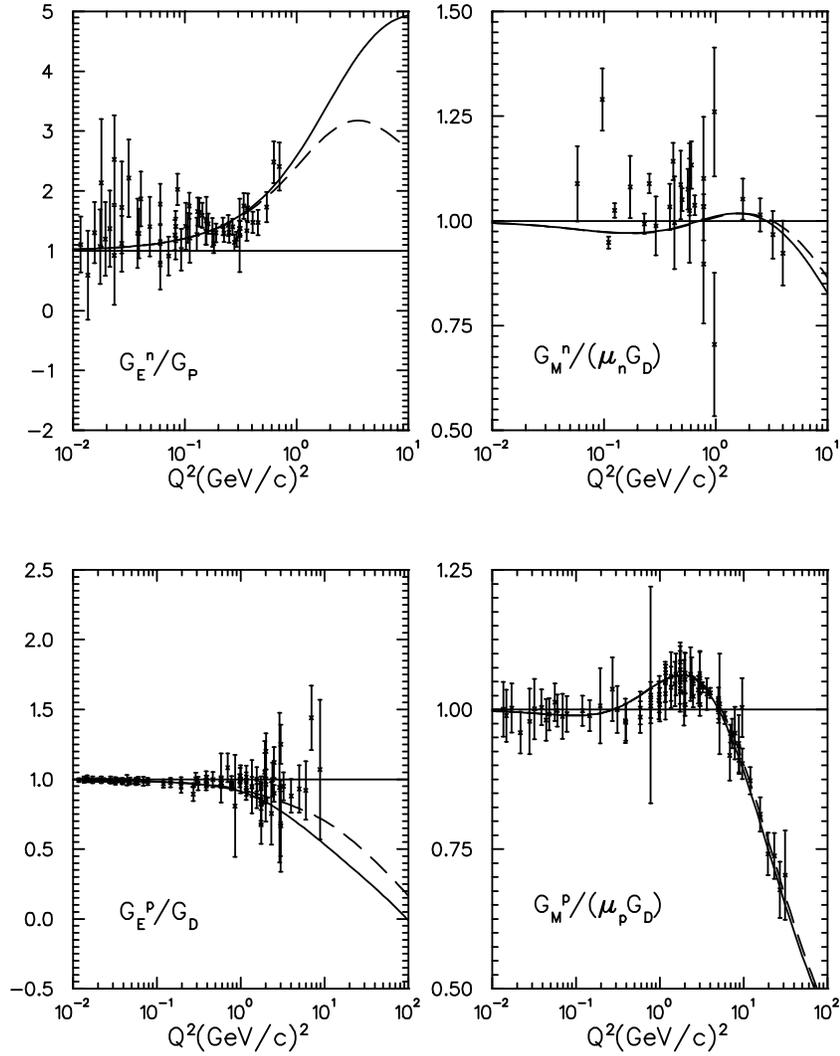}} 
\caption
{The Sachs form factors of neutron and proton as functions of $Q^2$, 
  normalized to the dipole ($G_D$) and ``Platchkov'' ($G_P$) form 
  factors defined in Eq.~(\ref{FF9}). The solid and dashed lines 
  are obtained from a fit to the world data, based on dispersion 
  relations. See Ref.~\protect\cite{Ham96} for more details. 
} 
\label{FF.fig5} 
\end{figure} 

In the case of the proton, the Rosenbluth plot was evaluated up to 
$Q^2=8.8$~GeV$^2$ at SLAC~\cite{Bos92}. The results are shown 
in Fig.~\ref{FF.fig5}. Additional and more precise information can be 
obtained at the new electron accelerators by double-polarization 
experiments, in particular by target polarization $\vec{p}(\vec{e}, 
e')p$ and recoil polarization, $p(\vec{e}, e')\vec{p}$. The asymmetry 
$A$ measured by such an experiment is given by~\cite{Arn81} 
\begin{equation} 
  A = - P_e \frac{\sqrt{2\tau\varepsilon(1-\varepsilon)}\  
G_EG_MP_x+\tau\sqrt{1-\varepsilon^2}\ G_M^2\ P_z} 
{\varepsilon\ G_E^2+\tau\ G_M^2}\ , 
\end{equation} 
where $P_e$ is the (longitudinal) polarization of the incident 
electron, and $P_x$ and $P_z$ are the transverse and longitudinal 
polarization components of the nucleon as defined in Eq.~(\ref{KIN9}). 
In particular we find that the longitudinal-transverse interference 
term, appearing if the nucleon is polarized perpendicularly (sideways) 
to $\vec{q}$, will be proportional to $G_EG_M$, while the 
transverse-transverse interference term, appearing for polarization in 
the $\vec{q}$ direction, will be proportional to $G_M^2$. The ratio of 
both measurements then determines $G_E/G_M$ with high precision, 
because most normalization and efficiency factors will cancel.

Within the large error bars of the experiments, the older data 
followed surprisingly close the so-called ``dipole fit'' for the Sachs 
form factors, 
\begin{eqnarray} 
G_E^p & = & G_M^p/\mu_p = G_M^n/\mu_n = (1+Q^2/M_V^2)^{-2}:= G_D\nonumber \\ 
G_E^n/\mu_n & = & \tau (1+Q^2/M_{V'}^2)^{-1}(1+Q^2/M_V^2)^{-2}:= G_P \ , 
\label{FF9} 
\end{eqnarray} 
with $\mu_p=2.79,\ \mu_n=-1.91$, $M_V=840$~MeV and $M_{V'}=790$~MeV. 
Since $\tau=Q^2/4m^2,\ G_E^n(0)$ vanishes, while $G_E^p(0)=1$ and the 
magnetic form factors approach the total magnetic moments for 
$Q^2\rightarrow 0$. In the asymptotic region $Q^2\rightarrow \infty$, 
all Sachs form factors should have a $Q^{-4}$ behavior according to 
perturbative QCD. Inverting Eq.~(\ref{FF6}) we also find the 
asymptotic behavior of the Dirac form factors as required by pQCD, 
$F_1\rightarrow Q^{-4}$ and $F_2\rightarrow Q^{-6}$. 
 
Already the SLAC experiments showed, however, that $G_M^p/\mu_p G_D$
falls much below unity at the higher momentum transfers~\cite{Arn86},
reaching values of about 0.65 at $Q^2=20$~(GeV/c)$^2$. For the reason
pointed out before, $G_E^p$ was not well determined by these
experiments. This situation has changed dramatically by the recent
results from Jefferson Lab, which were obtained by scattering
polarized electrons in coincidence with the polarization of the
recoiling protons~\cite{JLff}. In this way it was possible to separate
the form factors up to $Q^2=3.5$(GeV/c)$^2$, where $G_E^p/G_M^p$
reaches the surprisingly low value of about 0.55, i.e.  $G_E^p$ falls
below $G_D$ even faster than $G_M^p$.
 
\begin{figure}[h] 
\centerline{\epsfxsize=10cm \epsfbox{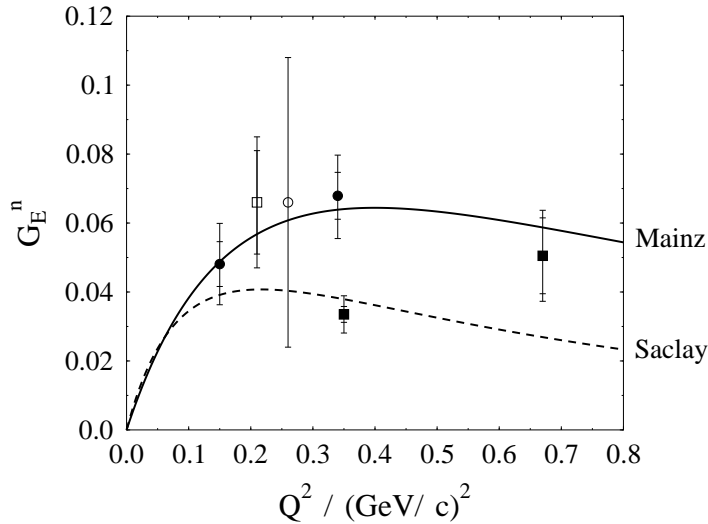}} 
\vspace{-0.5cm} 
\caption 
{The electric Sachs form factor for the neutron, $G_E^n$, as function
  of $Q^2$. The dashed line follows the Saclay data, the full line
  fits the new double-polarization experiments at Mainz. See
  Ref.~\protect\cite{Sch99} and text for details.  }
\label{FF.fig6} 
\end{figure} 

However, the situation is even more complex in the case of the
neutron. The only exact information used to be the electric neutron
radius, $\langle r^2\rangle_E^n\approx -0.11$~fm$^2$, which was
obtained by scattering low energy neutrons off a $^{208}Pb$
target~\cite{Kop95}.  Since there is no free neutron target, electron
scattering data have to be obtained from light nuclei such as $^2$H or
$^3$He making appropriate corrections for binding effects. This is a
particularly difficult task for $G_E^n$, because it is smaller than
the other form factors by a factor 10-20.  In the past, results were
obtained by either deuteron breakup in quasifree (neutron)
kinematics~\cite{Bar73} or elastic scattering off the
deuteron~\cite{Pla90}, assuming that all other form factors and wave
function corrections were well under control. Though the data reached
a remarkable statistical accuracy, large systematical errors remained,
particularly with regard to the nucleon-nucleon potential.  While it
had been pointed out long ago that double-polarization experiments
should be much less model-dependent, such data were only taken very
recently~\cite{Sch99}. As shown in Fig.~\ref{FF.fig6} the electric
form factor of the neutron seems to be much larger than previously
thought of. With the exception of the $^3$He point at $Q^2\approx
0.35$~(GeV/c)$^2$, the new data follow the full line (``Mainz fit'')
as opposed to the dashed line (``Saclay fit'', obtained from elastic
$ed$ scattering). It is remarkable that the $^2$H data point at the
lowest $Q^2$ has moved upward by nearly a factor of 2 by taking
account of final state interactions~\cite{Are87}, while these
corrections are only at the percent level for the higher $Q^2$.  This
observation is at variance with the earlier assumption that final
state interactions would not play any role in this kind of experiment.
In view of this lesson from the deuteron it may be assumed that also
the lowest $^3$He data point will move once a complete calculation of
final state and meson exchange effects exists.

The following Fig.~\ref{FF.fig7} compares the neutron charge density
obtained by Fourier transforming the older and the more recent data.
Both results are in qualitative agreement with our expectation that
the neutron charge density should have a positive core surrounded by a
negative cloud~\cite{LuT98}.  The remarkable facts are, however, that
the new data lead to a lower zero-crossing at r=0.7~fm in comparison
with the older results (r=0.9~fm), and that both maximum and minimum
become more pronounced. If one naively interprets the total negative
charge as the pion cloud, one finds a probability of about 60~\% that
the neutron has a proton core surrounded by a $\pi^-$ cloud. Such an
idea is quite natural for models of pions and nucleons, in particular
for chiral bag models. It is interesting to note that a similar
density is also predicted by the constituent quark model. The
hyperfine interaction leading to the $\Delta$-nucleon mass splitting
predicts, at the same time, a stronger repulsion of quarks with equal
flavor.  Therefore the two $d$ quarks with total charge $-2/3$ will
move to the bag surface while the up quark goes to the center.
 
\begin{figure}[h] 
\centerline{\epsfxsize=8cm \epsfbox{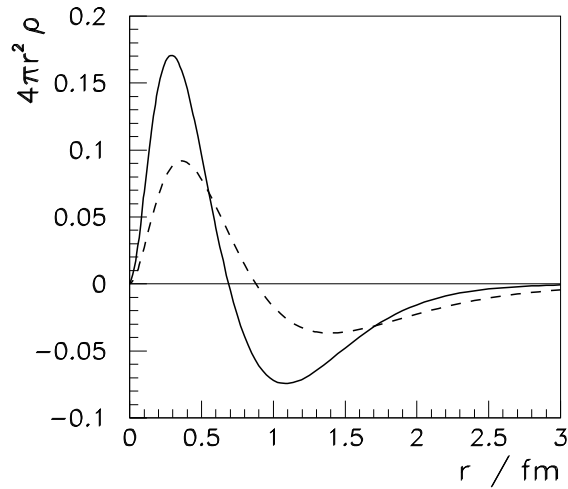}} 
\caption 
{The density distribution of the neutron, $\rho_E^n=\rho$, as function
  of the radius $r$. The two lines are the Fourier transforms of the
  corresponding fits in Fig.~\ref{FF.fig6}.  Results from
  Ref.~\protect\cite{Sch99a}.  }
\label{FF.fig7} 
\end{figure} 

\begin{table}[htpb] 
\begin{center} 
\caption 
{The proton charge radius $r_E^p=r_E$ derived from various experiments 
  and from a fit based on dispersion theory. Also shown is the radius 
  of an equivalent homogeneous sphere ($R_{eq})$ and the volume Vol 
  of that sphere. See text for details.} 
\label{FF.tab1} 
\vspace{0.5cm} 
\begin{tabular}{|l|r|r|r|} 
\hline 
                     & $r_E/$fm   &  $R_{eq}/$fm   &  Vol/fm$^3$ \\ 
\hline 
Stanford~\cite{Han63}         & 0.81       &  1.05          &   4.85 \\ 
disp. theory~\cite{Ham96}      & 0.85       &  1.10          &   5.58 \\ 
Mainz~\cite{Sim80}             & 0.86       &  1.11          &   5.73 \\ 
opt. and rf. exp.~\cite{Kar97} & 0.92       &  1.19          &   7.06 \\ 
\hline 
\end{tabular} 
\end{center} 
\end{table}

A final remark is in order concerning the proton radius. The
experimental situation for $r_E^p$ is shown in Table~\ref{FF.tab1}.
The large data spread for this very elementary quantity is truly
surprising. The recent optical and radio frequency experiments had in
mind, of course, to search for the limits of quantum electrodynamics
by measuring Lamb shifts and hyperfine structures. In spite of an
astounding accuracy of about 12 decimals, the analysis was stopped at
about the 7th decimal by the existing uncertainties in the proton
radius.  If all deviations from theory are attributed to size effects,
considerably larger radii are obtained than in the case of electron
scattering.

The size of the nucleon is not just an academical question, but of 
tremendous consequence for our understanding of hadronic matter. The 
Table also shows the radius of an equivalent, homogeneous charge 
distribution, $R_{eq}=(\frac{5}{3} r_E^2)^{1/2}$, and the resulting 
``volume'' of a nucleon. Obviously the volume grows by nearly 50~\% by 
going from the Stanford value to the more recent results. Hence 
nucleons in a nucleus may get into a very uncomfortable environment: 
they need more space than is actually available. This situation is of 
course quite different from most models of nuclei and nuclear matter, 
which are based on effective interactions between point particles. 
 
Note added in proof: In a recent paper Rosenfelder pointed out that 
Coulomb corrections will increase the proton radius, as measured by 
electron scattering, to $r_E=(0.880\pm0.015)$~fm, with an error bar 
depending on the fit strategy~\cite{Ros99}.

\section{STRANGENESS} 
The strangeness content of the nucleon manifests itself by matrix 
elements $<N\mid\bar{s}\ \Gamma\ s\mid N>$, with $\Gamma$ any of the 5 
Dirac structures of Section 3. Though observation of these matrix 
elements is necessarily proof for the existence of $s$ quarks in the 
nucleon, the strength of the 5 matrix elements may well be different. 
Since there exists no net strangeness in the nucleon, these observables 
open, in principle, a clear window on sea degrees of freedom. 
Information on the strange quark content comes essentially from three 
sources: 
\begin{enumerate} 
\item[(1)] Deep inelastic lepton scattering. \\ The experiments 
  clearly indicate a break-down of the Ellis-Jaffe sum rule based on 
  SU(3) symmetry~\cite{Ell74}. Such experiments have led to the 
  so-called ``spin crisis'' of the nucleon, which was eventually 
  explained by sea quark and gluon contributions to the spin of the 
  nucleon. The observed symmetry breaking is proportional to the axial 
  vector current carried by the $s$ quarks~\cite{Ell95}. 
\item[(2)] Pion-nucleon scattering. \\ Dispersion analysis allows one to 
  extrapolate $\pi N$ scattering to the (unphysical) Cheng-Dashen 
  point at $s=u$ and $t=2m_{\pi}^2$. The scattering amplitude at this 
  point is essentially given by the $\sigma$ term~\cite{Gas91}, 
\begin{equation} 
  \sigma_{\pi N} = \frac{m_u+m_d}{2} <N\mid\bar{u}u+\bar{d}d\mid N>\ . 
\end{equation} 
In combination with similar information on $KN$ scattering and 
approximate SU(3), the scalar $\bar{s}s$ condensate can be determined. 
The size of this effect is, unfortunately, not well established. The 
prediction of ChPT is~\cite{Bor97} 
\begin{equation} 
  \sigma_{\pi N} = 58.3\ (1-0.56+0.33)\ {\mbox{MeV}} = 45\ {\mbox{MeV}}\ , 
\end{equation} 
the 3 terms in this equation indicating the (slow) convergence of the
perturbation series, while typical phenomenological analyses are in
the range of $\sigma_{\pi N}=(60\pm 20)$~MeV. It is obvious that this
uncertainty will also affect the value of the strangeness
contribution, which therefore carries large error bars,
\begin{equation} 
 y = \frac{2<\bar{s}s>}{<\bar{u}u+\bar{d}d>} = 0.21\pm 0.20\ . 
\end{equation} 
\item[(3)] Parity-violating lepton scattering. \\ The interest in this 
  experiment stems from the observation that the photon and the 
  $Z^0$ gauge boson couple differently to the vector currents of 
  the quarks. An interference term of the electromagnetic and the weak 
  neutral current is parity-violating (PV) and thus can be determined 
  by a PV asymmetry. This presents the opportunity to measure a third 
  form factor, in addition to the electromagnetic form factors of 
  neutron and proton. If these 3 form factors would be known to 
  sufficient precision, the density distributions of $u,\ d$ and $s$ 
  quarks could be determined~\cite{Mus92}. 
\end{enumerate} 
 
The strange vector current $<N'\mid\bar{s}\gamma_{\mu}s\mid N>$ takes 
the general form of Eq.~(\ref{FF4}) with Dirac $(F_1^s)$ and Pauli 
$(F_2^s)$ strangeness form factors. Since the nucleon has no net 
strangeness, $F_1^s(Q^2=0)=0$. It follows from Eq.~(\ref{FF6}) that 
\begin{eqnarray} 
 G_E^s (Q^2) & = & -\frac{1}{6} Q^2 <r^2>_E^s+[Q^4]\ ,\nonumber \\ 
 G_M^s (Q^2) & = & \mu^s+[Q^2]\ , 
\label{STR7} 
\end{eqnarray} 
with $<r^2>_E^s$ the square of the electric $rms$ radius and
$\mu^s=\kappa^s=F_2^s(0)$ the (anomalous) magnetic moment due to the
strange quark sea. Instead of the radius one often finds the
dimensionless quantity $\varrho^s=dG/d\tau$, the derivative of a
particular form factor with regard to the quantity $\tau$, which is
related to
\begin{equation} 
 <r^2>^s = -0.066\ {\mbox{fm}}^2\ \varrho^s\ . 
\end{equation} 

The new information from parity-violating $\vec{e}+N\rightarrow e'+N'$ 
can be obtained from the asymmetry 
$A=(d\sigma^+-d\sigma^-)/(d\sigma^++d\sigma^-)$, where $d\sigma^+$ and 
$d\sigma^-$ denote the cross sections for positive and negative 
helicities of the incident electron. While such an asymmetry must 
vanish in the purely electromagnetic case, it can appear by an 
interference between the leading electromagnetic and the much smaller 
(parity violating) weak interaction.  Of course, the leading term is 
obtained by the absolute square of the amplitudes for photon exchange, 
resulting in a contribution $O(e^4)$, while the subleading term is 
given by the interference of photon exchange and $Z^{\circ}$ exchange, 
which is $O(e^2 G_F^2)$, with $G_F$ Fermi's constant of weak 
interactions.  While the photon couples only via the vector current, 
the $Z^{\circ}$ can couple both to vector and axial currents. The 
interesting, parity violating interference term appears if the 
$Z^{\circ}$ couples with the vector current to the nucleon and with 
the axial current to the electron or vice versa. 
 
Table~\ref{STRtab2} shows, in standard notation, the vertices for the 
coupling of some leptons and quarks to photon and $Z^{\circ}$, 
\begin{table} 
\begin{center} 
\caption 
{The vertices for the couplings of photons and $Z^0$ gauge bosons to 
  electrons (e), neutrinos ($\nu$), and quarks (u,d,s).} 
\label{STRtab2} 
\vspace{0.5cm} 
\begin{tabular}{|l|l|l|} 
\hline & &\\  & photon & $Z^{\circ}$ gauge boson \\ 
\hline & & \\ $e$ & $-ie\gamma_{\mu}$ & $ 
  ig'\gamma_{\mu}(1-4s^2-\gamma_5)$ \\ & & \\ 
 $\nu$ & 0 & $ 
  -ig'\gamma_{\mu}(1-\gamma_5)$ \\ & & \\ 
$u$ &  $+\frac{2}{3}ie\gamma_{\mu}$ & $ 
  -ig'\gamma_{\mu}(1-\frac{8}{3}s^2-\gamma_5)$ \\ 
 & & \\ $d,s$ & $-\frac{1}{3}ie\gamma_{\mu}$ 
 & $ ig'\gamma_{\mu}(1-\frac{4}{3}s^2-\gamma_5)$,  \\ 
 & & \\ \hline 
\end{tabular} 
\end{center} 
\end{table} 
where $g'=e/4sc$, $s=\sin\theta_W$, $c=\cos\theta_W$, and $\theta_W$
is the Weinberg angle, given by $\sin^2\theta_W=0.2319\pm 0.0005$. We
observe that the coupling of the electron to the $Z^{\circ}$ is
dominated by the axial vector, because the vector part is suppressed
by $4\sin^2\theta_W\approx 1$. By the same fact the vector currents of
the quarks couple quite differently to photons and $Z^{\circ}$ bosons,
in particular the ratio of $u$ to $d$ or $s$ quark couplings reverses
from -2 about $-\frac{1}{2}$ if going from electromagnetic to weak
neutral interactions.
 
The experimental information on strangeness is given by the 
asymmetry, which in the case of the nucleon takes the form 
\begin{eqnarray} 
  {\cal A} & = & 
  \frac{d\sigma^+-d\sigma^-}{d\sigma^++d\sigma^-} \nonumber \\ 
  & = &  \frac{\frac{G_E\tilde{G}_E+\tau G_M\tilde{G}_M}{1+\tau} 
        +2\tau G_M\tilde{G}_M\tan^2\frac{\theta}{2} + ...\ 
       (1-4s^2)G_M\tilde{G}_A } 
       {\frac{G_E^2+\tau G_M^2}{1+\tau} 
       +2\tau G_M^2\tan^2\frac{\theta}{2}} \nonumber \\ 
  & = & {\cal A}^E (\tilde{G}_E) + {\cal A}^M (\tilde{G}_M) + 
        {\cal A}^A (\tilde{G}_A) \ . 
\end{eqnarray} 
Though the 3 form factors $\tilde{G}_E,\ \tilde{G}_M,\ \tilde{G}_A$ 
can in principle be separated by a super Rosenbluth plot, definite 
results will take some time. The total asymmetry in a typical 
experiment is ${\cal A} \approx 10^{-4} Q^2/$~(GeV/c)$^2$, and only a 
small fraction of ${\cal A}$ is due to the expected effects of the 
strange quarks. According to Table~\ref{STRtab2} these effects can be 
obtained by the quark currents 
\begin{equation} 
  J_{\mu}^{(\gamma)} = \sum e_q\ \bar{q}\gamma_{\mu}q \ \ , \ \ 
  J_{\mu}^{(Z_0)} = \sum \tilde{e}_q\ \bar{q}\gamma_{\mu}q\ , 
\end{equation} 
with $e_u=2/3$, $e_d=e_s=-1/3$, $\tilde{e}_u=-1+8s^2/3$ and 
$\tilde{e}_d=\tilde{e}_s=1-4s^2/3$, where $s^2=\sin^2\theta_W$. The 
matrix element of these quark currents between nucleon states can be 
parametrized by form factors describing the quark structure, e.g. 
\begin{equation} 
  \langle p'\mid \bar{s}\gamma_{\mu}s\mid p\rangle = 
  G_s(Q^2)\bar{u}_{p'}\gamma_{\mu}u_p +\ {\mbox{magnetic\ terms}}\ . 
\end{equation} 
The sum of the $u$, $d$, and $s$ quark contributions must equal the 
form factor of the nucleon, 
\begin{eqnarray} 
  G^p & = & \frac{2}{3} G_u-\frac{1}{3}\ (G_d+G_s)\ , \nonumber \\ 
  G^n & = & \frac{2}{3} G_d-\frac{1}{3}\ (G_u+G_s)\ , \nonumber \\ 
  \tilde{G}^p & = & \left (-1+\frac{8}{3}s^2\right ) G_u+ 
  \left (1-\frac{4}{3}s^2\right ) (G_d+G_s)\ . 
\label{STR22} 
\end{eqnarray} 
In these equations, $G_{u/d/s}$ are the quark distributions in the 
proton, and those of the neutron have been assumed to follow from 
isospin symmetry. If the 3 form factors on the $lhs$ of 
Eq.~(\ref{STR22}) have been measured, the strange quark contribution 
can be determined from 
\begin{equation} 
  \tilde{G}^{p} = -(1-4\sin^2\theta_W) 
  G^{p}+G^{n}+G^s\ . 
\end{equation} 
A particularly simple formula may be obtained for PV 
scattering~\cite{Pre78} off $^4$He. Since this nucleus has spin zero, 
there exists only a charge monopole form factor. Furthermore $^4$He is 
well described by an isoscalar system of nucleons having the same 
spatial wave functions.  Under these assumptions the asymmetry may be 
cast into the form 
\begin{equation} 
  {\cal A}\ (^4{\mbox{He}}) = \frac{G_FQ^2}{\pi\sqrt{2}\alpha_{fs}} 
  \left (\sin^2\theta_W + \frac{G_E^s}{2(G_E^p+G_E^n)}\right )\ . 
\end{equation} 
With $G_F$ the Fermi constant and $\alpha_{fs}$ the fine structure
constant, the factor in front of the bracket is about $4\cdot10^{-4}
Q^2/$~(GeV/c)$^2$, and with the value of $\theta_W$ the ``non
strange'' asymmetry is about $10^{-4} Q^2/$~(GeV/c)$^2$, which sets
the scale for this difficult experiment. While earlier experiments on
PV electron scattering~\cite{Pre78} were performed in order to
determine $\theta_W$, which of course required that $G_E^s$ be
negligible, the Weinberg angle is now known to 3 digits and today the
motivation is to determine the strange quark contribution.
 
The simplest model for the strangeness contribution is, say, a proton 
that part of the time contains a strange pair, 
\begin{equation} 
 \mid p\rangle = \mid u^2d\rangle+\mid u^2ds\bar{s}\rangle + ...\ , 
\label{STR25} 
\end{equation} 
with the ellipse standing for $u$ and $d$ pairs and higher
configurations. As long as $s$ and $\bar{s}$ quarks have the same
spatial wave function, their charges cannot be seen by the electron.
In order to separate the quarks in space however, the wave functions
have to be correlated, the simplest long-range correlation being the
clustering of the second component in Eq.~(\ref{STR25}) in the form of
$\Lambda(uds)\otimes K^+(u\bar{s})$. This model will therefore
predict, as contribution of the strange sea, a positively charged
cloud $(K^+)$ and a negative core (the neutral $\Lambda$ relative to
the charged $p$)~\cite{Mus97}. As a result both the anomalous magnetic
moment of the proton, $\kappa_p$, and the value of $\langle
r^2\rangle_E^p$ will be increased. Since the $s$ quark has negative
charge, this model predicts $\mu^s=\kappa^s<0$ and $\langle
r^2\rangle_E^s<0$ for the quantities introduced in Eq.~(\ref{STR7}).
 
A second model is based on dispersion relations, which tend to predict 
a strong contribution of the $\Phi$(1020) in order to combine with the 
$\omega$(780) to an approximate dipole form of the isoscalar form 
factors~\cite{Jaf89}. Since the $\Phi$ is practically an $s\bar{s}$ 
configuration, its appearance is related with strangeness in the 
nucleon. Other calculations have been performed in Skyrme, chiral 
quark-soliton and constituent quark models, and in the framework of 
lattice QCD and ChPT. Such calculations generally result in negative 
values for $\mu^s$ with a range of $-.3 \gtrsim \mu^s\gtrsim -.7$, 
while $\langle r^2\rangle_E^s\approx0.15$~fm$^2$ in dispersion models 
($\Phi$ poles) and $0\gtrsim \langle r^2\rangle_E^s\gtrsim 
-0.15$~fm$^2$ for $K$ loops. 
 
The recent results of the SAMPLE experiment at MIT/Bates and of 
HAP-PEX at Jefferson Lab came as a big surprise: The $s$ quark 
contribution is much smaller than predicted, and in fact even 
compatible with zero.  The SAMPLE experiment measured essentially 
$G^s_M$, which came out positive though with large error 
bars~\cite{Mue97}. Extrapolating to $G_M^s(0)=\mu^s$, Hemmert et 
al.~\cite{Hem98a} obtained $0.03<\mu_p^s<0.18$ by use of the slope of 
$G_M^s (Q^2)$ as predicted from HBChPT (note that this theory cannot 
predict $\mu^s$ itself, because of an unknown low energy constant). 
The HAPPEX collaboraton obtained a raw asymmetry ${\cal 
  A}=-5.64\pm0.75$~ppm.  Since most of this asymmetry was expected on 
the basis of $u$ and $d$ quarks, only a small fraction remained as 
possible $s$ quark contribution, leading to the result~\cite{Ani98} 
\begin{equation} 
  G_E^s+0.39 G_M^s = 0.023\pm 0.034\pm 0.022 \pm 0.026 
\label{STR26} 
\end{equation} 
at $Q^2=0.48$ (GeV/c)$^2$. The error bars in Eq.~(\ref{STR26}) denote,
in order of appearance, the statistical and systematical uncertainties
as well as the errors due to our bad knowledge of the neutron from
factor $G_E^n$ at that momentum transfer. The result is again positive
though with large error bars, and taken at face value it rules out
most theoretical predictions. A selection of these predictions can be
found in Ref.~\cite{Pit99}. Contrary to earlier lattice QCD
predictions, a recent lattice calculation finds small negative values
$G_M^s(0) = -0.16\pm0.18$, which could even shift to more positive
values because of systematic errors~\cite{Lei99}. From a comparison of
recent data obtained for proton and deuteron targets, it has been
suspected that the hadronic radiative corrections to the axial form
factor are not yet under control. In view of the importance of this
topic, more and new experiments on the strange form factor are
underway~\cite{Bec91}.

 
\section{COMPTON SCATTERING} 
 
The polarizability measures the response of a particle to a 
quasistatic electromagnetic field. In particular the energy is 
generally lowered by 
\begin{equation} 
  \Delta E = -\frac{1}{2}\alpha\vec{E}\ ^2-\frac{1}{2}\beta\vec{H}\ ^2 
  \ , 
\end{equation} 
where $\vec{E}$ and $\vec{H}$ are the electric and magnetic fields, 
and $\alpha$ and $\beta$ the electric and magnetic polarizabilities. In 
the case of a macroscopic system with N atoms per volume, the 
polarizabilities are related to the dielectric constant $\varepsilon$ 
and the magnetic permeability $\mu$ by 
\begin{equation} 
 \varepsilon = 1-N\alpha\ \ ,\ \ \mu = 1-N\beta\ . 
\end{equation} 
The electric polarizability of a metal sphere is essentially given by 
its volume, it scales with the third power of the radius. In the case 
of a dielectric sphere an additional factor 
$(\varepsilon-1)/(\varepsilon+2)$ appears, which reduces the 
polarizability by orders of magnitude, because $\varepsilon$ is close 
to unity. The same is true for the nucleon. If we divide its 
polarizability by the volume $V$, we obtain 
\begin{equation} 
  \frac{\alpha}{V} \approx 
  \frac{10^{-3}{\mbox{fm}}^3}{\frac{4}{3}\pi{\mbox{fm}}^3} \approx 
  2\cdot10^{-4}\ , 
\end{equation} 
i.e. the nucleon is a very rigid object. It is held together by strong
interactions, and the applied electromagnetic field cannot easily
deform the charge distribution.  Of course the nucleon cannot be
polarized by putting it between two condensator plates. Instead its
polarizability can be measured by Compton scattering: The incoming
photon deforms the nucleon, and by measuring the energy and angular
distributions of the outgoing photon one can determine the
polarizability.
 
In nonrelativistic quantum mechanics the electric polarizability is 
given by 
\begin{equation} 
  4\pi\alpha = 2\hbar^2\sum_{n>0}\frac{\mid\langle 
    n\mid\hat{D}_z\mid0\rangle\mid^2}{E_n-E_0}\ , 
\label{CS4} 
\end{equation} 
where $\hat{D}_z=e\hat{z}$ is the dipole operator and 
$e^2/4\pi\approx1/137.$ Since all excitation energies $E_n-E_0$ are 
positive and only the modulus of the transition matrix element 
$\langle n\mid\hat{D}_z\mid0\rangle$ enters, $\alpha$ has to be 
positive in a nonrelativistic model.
 
Here is a simple prototype problem for a polarizable 
system~\cite{NatBo}. A nonrelativistic particle with mass $M$ and 
charge $Q$ is held by a harmonic oscillator potential with Hooke's 
constant $C=M\omega_0^2$.  If we apply an external electrical field 
$\vec{E}$, the Hamiltonian is 
\begin{equation} 
  H = \frac{\vec{p}\ ^2}{2M} + \frac{M\omega_0^2}{2}\vec{r}\ ^2 + Q 
  \vec{E}\cdot\vec{r}\ , 
\end{equation} 
which can be cast into the form 
\begin{equation} 
  H = \frac{\vec{p}\ ^2}{2M} + \frac{M\omega_0^2}{2}\left 
    (\vec{r}+\frac{Q}{M\omega_0^2}\vec{E}\right 
  )^2-\frac{1}{2}\frac{Q^2}{M\omega_0^2}\vec{E}^2\ . 
\end{equation} 
The result is 
\begin{enumerate} 
\item[(i)] a shift in space, 
  $\Delta\vec{r}=\frac{Q}{M\omega_0^2}\vec{E}$, leading to an induced 
  dipole moment $\vec{d} = Q \Delta\vec{r}:=\alpha\vec{E}$, and 
\item[(ii)] a shift in energy, $\Delta E=-\frac{1}{2}\alpha\vec{E}^2$, 
  with $\alpha=\frac{Q^2}{M\omega_0^2}$. 
\end{enumerate} 
 
In view of several misrepresentations in the literature, we 
stress the point that these two definitions of $\alpha$, via induced 
dipole moment or energy shift, should lead to the same value. 
 
A more generic model involves two particles (masses $M_1$ and $M_2$,
charges $Q_1$ and $Q_2$), held together with a spring constant
$C=\mu\omega_0^2$, where $\mu$ is the reduced mass. An external field
$\vec{E}$ induces both an intrinsic dipole moment (expressed in terms
of the relative coordinate) and an acceleration of the center of mass.
According to classical antenna theory, the scattering amplitude
$f(\omega)$ is proportional to the acceleration of the induced dipole
moments. The final result is~\cite{NatBo}
\begin{eqnarray} 
  f(\omega) & = & -\frac{Q^2}{M}+\frac{1}{\mu(\omega_0^2-\omega^2)}\left 
    (\frac{Q_1M_2-Q_2M_1}{M}\right )^2\omega^2 \nonumber \\ &  = & 
  -\frac{Q^2}{M}+4\pi\alpha(\omega)\omega^2\ . 
\label{CS7} 
\end{eqnarray} 
In the limit of $\omega\rightarrow 0$, the scattering amplitude
reduces to the Thomson term depending only on the total charge $Q$ and
the total mass $M$ of the system. It is the essence of more refined
``low energy theorems'' (LET) that only such global properties should
be visible in that limit. Since the cross section $d\sigma/d\Omega\sim
\mid f(\omega)\mid^2$, the internal structure shows up first at
$O(\omega^2)$, as interference of the Thomson term with the second
term in Eq.~(\ref{CS7}). In the case of a globally neutral system
(e.g. a neutral atom, a neutron or a $\pi^0$), the Thomson term
vanishes and the cross section starts at $O(\omega^4)$.  This is the
familiar case of Rayleigh scattering leading to the blue sky, because
most gases absorb in the ultra-violet, $\omega_0^2\gg\omega^2$, with
$\omega$ a frequency of visible light. If $\omega$ increases further,
it approaches a singularity in Eq.~(\ref{CS7}), which is of course
avoided by appropriate friction terms, i.e. by a width $\Gamma_0$ of
the resonance at $\omega_0$.
 
Compton scattering off the proton is, of course, technically much more
complicated than the nonrelativistic model above. The reasons are
relativity and the spin degrees of freedom. By use of Lorentz and
gauge invariance, crossing symmetry, parity and time reversal
invariance, the general Compton amplitude takes the form~\cite{Lvo97}
\begin{equation} 
  T = \varepsilon'^{\ast}_{\mu}\varepsilon_{\nu}\sum_{i=1}^6 {\cal 
    O}_i^{\mu\nu} \tilde{A}_i(s,t)\ , 
\end{equation} 
where ${\cal O}_i^{\mu\nu}$ are Lorentz tensors constructed from 
kinematical variables and $\gamma$ matrices, and $\tilde{A}_i$ are 
Lorentz scalars. In the $cm$ frame, these Lorentz structures can be 
reduced to Pauli matrices combined with unit vectors in the directions 
of the initial $(\hat{k})$ and final $(\hat{k}')$ photons, which 
yields the result~\cite{Lvo97,Hem98a} 
\begin{eqnarray} 
\label{CS14} 
T&=&A_1(\omega , t)\vec{\epsilon}\ '^{\ast} \cdot \vec{\epsilon}+ 
    A_2(\omega , t)\vec{\epsilon}\ {'}^{\ast} \cdot \hat{k}\vec{\epsilon} 
        \cdot \hat{k}{'}\nonumber \\ 
  && +  iA_3(\omega , t)\vec{\sigma} \cdot 
        (\vec{\epsilon}\ {'}^{\ast} \times \vec{\epsilon})+ 
      iA_4(\omega , t)\vec{\sigma} \cdot (\hat{k}{'} \times \hat{k}) 
        \vec{\epsilon}\ {'}^{\ast} \cdot \vec{\epsilon} \nonumber \\ 
  && +  iA_5(\omega , t)\vec{\sigma} \cdot \left[(\vec{\epsilon}\ {'}^{\ast} 
       \times \hat{k})\vec{\epsilon} \cdot \hat{k}{'}-(\vec{\epsilon} 
       \times \hat{k}')\vec{\epsilon}\ {'}^{\ast}\cdot \hat{k} \right ]  \nonumber \\ 
  && +  iA_6(\omega , t)\vec{\sigma} \cdot \left[(\vec{\epsilon}\ {'}^{\ast} 
       \times \hat{k}')\vec{\epsilon} \cdot \hat{k}{'}-(\vec{\epsilon} 
       \times \hat{k})\vec{\epsilon}\ {'}^{\ast}\cdot \hat{k} \right ]\ , 
\end{eqnarray} 
with $\hat{\epsilon}$ and $\hat{\epsilon}'$ describing the 
polarization of the photon in the initial and final states, and 
$\vec{\sigma}$ the spin of the nucleon. 
 
The low energy theorem predicts the following threshold behavior for 
the proton amplitudes~\cite{Hem98a}: 
 
\begin{eqnarray} 
\label{CS15} 
A_1 & = & -\frac{e^2}{m} + 4\pi(\alpha+\beta \cos\theta)\omega^2 
- \frac{e ^2}{4m^3}(1-\cos\theta)\omega^2 + \ ...\ 
,\nonumber \\ 
 A_2 & = & \frac{e^2}{m}\omega-4\pi\beta\omega^2 +\ ...\ , 
\nonumber \\ 
 A_3 & = & [(1+2\kappa)(1-\cos\theta) - 
\kappa^2\cos\theta] \frac{e ^2\omega}{2m ^2} - 
\frac{(2\kappa+1)e ^2}{8m ^4} \cos\theta~\omega^3 \nonumber \\ 
&& + 4\pi [\gamma_1-(\gamma_2+2\gamma_4)\cos\theta] 
\omega^3+\ ...\ ,\nonumber \\ 
A_4 & = & -\frac{(1+\kappa)^2e^2}{2m^2}\omega+4\pi\gamma_2\omega^3 +\ ... \ ,
\nonumber \\ 
A_5 & = & \frac{(1+\kappa)^2e^2}{2m^2}\omega+4\pi\gamma_4\omega^3 +\ ...\ ,
 \nonumber \\ 
A_6 & = &-\frac{(1+\kappa)^2e^2}{2m^2}\omega+4\pi\gamma_4\omega^3 +\ ... \ .
\end{eqnarray} 
In the expansion for $A_1$ we recover the previously discussed low
energy theorem for forward scattering. In addition to $\alpha$,
however, also the magnetic polarizability $\beta$ appears. Since
$\alpha$ and $\beta$ enter differently in $A_1$ and $A_2$, they can be
determined separately by Compton scattering. The amplitudes $A_1$ and
$A_2$ are typical for a scalar (or pseudoscalar) particle, and for
this reason we call $\alpha$ and $\beta$ the scalar polarizabilities.
Since the nucleon has a spin, there appear 4 more amplitudes, $A_3$ to
$A_6$, whose leading terms , ${\cal O}(\omega)$, are related to the
magnetic moment $\mu=1+\kappa$. The subleading terms, ${\cal
  O}(\omega^3)$, define 4 new polarizabilities $\gamma_1$ to
$\gamma_4$, the spin or vector polarizabilities of the nucleon. We
recall that the differential cross section for small $\omega$ is
dominated by the Thomson term and that the polarizabilities $\alpha$
and $\beta$ appear in the cross section at ${\cal O}(\omega^2)$ via
the interference of Thomson and Rayleigh scattering. In addition,
however, also the spin-dependent amplitudes contribute at ${\cal
  O}(\omega^2)$ for unpolarized Compton scattering, because without
polarization the terms with and without the $\vec{\sigma}$ matrices
add incoherently in the cross section. For the same reason the spin
polarizabilities show up only at ${\cal O}(\omega^4)$, i.e. are
expected to be small and difficult to disentangle from other higher
order terms. It is therefore obvious that the 6 polarizabilities
cannot be determined from differential cross section measurements
only, but that polarization experiments are necessary, in particular
the scattering of circularly polarized photons off polarized protons.
 
In the following we shall again restrict the discussion to forward 
scattering, i.e. $\hat{k}'=\hat{k}$ or $\theta=0$. Due to the 
transversality condition 
$\hat{\epsilon}\cdot\hat{k}=\hat{\epsilon}'\cdot\hat{k}'$, only the 
amplitudes $A_1$ and $A_3$ contribute in that limit. With the notation 
$A_1(\omega,0)=f(\omega)$ and $A_3(\omega,0)=g(\omega)$, 
Eq.~(\ref{CS14}) can be cast into the form 
\begin{equation} 
  T (\omega, \theta = 0) = \hat{\epsilon}'^{\ast} \cdot \hat{\epsilon} 
  \ f(\omega) + i (\hat{\epsilon}'^{\ast} \times \hat{\epsilon}) \cdot 
  \vec{\sigma}\ g(\omega)\ . 
\end{equation} 
Due to the crossing symmetry, the non spin-flip amplitude $f(\omega)$
is an even function in $\omega$ and the spin-flip amplitude
$g(\omega)$ is odd. The 2 scattering amplitudes can be determined by
scattering circularly polarized photons (spin projection +1) off
nucleons polarized in the direction or opposite to the photon momentum
(spin projections +1/2 or -1/2), leading to intermediate states with
spin projection +3/2 or +1/2 respectively. Denoting the corresponding
scattering amplitudes by $T_{3/2}$ and $T_{1/2}$, we find
$f(\omega)=(T_{1/2}+T_{3/2})/2$ and $g(\omega)=(T_{1/2}-T_{3/2})/2$.
The optical theorem allows us to express the imaginary parts of $f$
and $g$ by the sum and difference of the helicity cross sections for
physically allowed values of $\omega$,
\begin{eqnarray} 
  {\mbox{Im}}\ f(\omega) & = & 
  \frac{\omega}{4\pi}\frac{\sigma_{1/2}+\sigma_{3/2}}{2} = 
  \frac{\omega}{4\pi}\sigma_{tot}(\omega) \nonumber \\ {\mbox{Im}}\ 
  g(\omega) & = & 
  \frac{\omega}{4\pi}\frac{\sigma_{1/2}-\sigma_{3/2}}{2} 
  =\frac{\omega}{4\pi} \Delta\sigma(\omega)\ . 
\end{eqnarray} 
We further assume that $f$ obeys a once-subtracted and $g$ an 
unsubtracted dispersion relation. Finally, we shall restrict the 
discussion to photon energies below pion threshold $\omega_0$, 
in which case the amplitudes are real and the dispersion relations can 
be cast into the form 
 
\begin{eqnarray} 
\label{CS18} 
4\pi\ f(\omega) & = & 4\pi\ f(0) + \frac{2\omega^2}{\pi} \int^{\infty} 
_{\omega_0}\frac{\sigma_{tot}(\omega')} {\omega'^2-\omega^2} 
d\omega'\ , 
\\ \nonumber 
4\pi\ g(\omega) & = & \frac{2\omega}{\pi} \int^{\infty} 
_{\omega_0}\frac{\Delta\sigma(\omega')}{\omega'(\omega'^2-\omega^2)} 
d\omega'\ , 
\end{eqnarray} 
 
which involves integrations from the physical threshold for pion 
production, $\omega_0$, to infinity.
 
Next we make use of the low-energy theorem~\cite{Low54}, which allows
us to express the low-energy behavior of $f(\omega)$ and $g(\omega)$
by a power series according to Eq.~(\ref{CS15}),
 
\begin{eqnarray} 
\label{CS19} 
4\pi\ f(\omega) & = & -\frac{e^2}{m} + 4 \pi (\alpha + \beta) \omega^2 + 
 [\omega^4] \ , \\ \nonumber 
4\pi\ g(\omega) & = & -\frac{2\pi e^2\kappa^2}{m^2} 
 + 4 \pi \gamma_0 \omega^3 + [\omega^5]\ . 
\end{eqnarray} 
 
If we compare Eqs.~(\ref{CS18}) and~(\ref{CS19}), we obtain a 
series of sum rules, in particular Baldin's sum rule~\cite{Bal60} 
 
\begin{equation} 
\alpha + \beta = \frac{1}{2\pi^2}\int^{\infty} 
_{\omega_0}\frac{\sigma_{tot}(\omega)}{\omega^2} 
d\omega \ , 
\label{CS20} 
\end{equation} 
the sum rule of Gerasimov, Drell and Hearn~\cite{Ger65}, 
 
\begin{equation} 
\kappa^2  = - \frac{2m^2}{\pi e^2}\int^{\infty} 
_{\omega_0}\frac{\sigma_{1/2}(\omega)- \sigma_{3/2}(\omega)}{\omega} 
d\omega \ , 
\label{CS21} 
\end{equation} 
and a value for the forward spin polarizability~\cite{Gel54}, 
 
\begin{equation} 
\gamma_0 = \frac{1}{4\pi^2} \int^{\infty} 
_{\omega_0}\frac{\sigma_{1/2}(\omega)- \sigma_{3/2}(\omega)}{\omega^3} 
d\omega \ . 
\label{CS22} 
\end{equation} 
 
Both the forward spin polarizability $\gamma_0$ and the GDH sum rule 
depend on the difference of the helicity cross sections, 
 
\begin{equation} 
\sigma_{1/2}-\sigma_{3/2} \sim |E_{0^+}|^2 - |M_{1^+}|^2 + 
E_{1^+}^{\ast} M_{1^+} + ...\ , 
\end{equation} 
 
i.e. are dominated by the difference of s-wave pion production 
(multipole $E_{0^+}$) and magnetic excitation of the $\Delta$(1232) 
resonance (multipole $M_{1^+}$). 
 
With the advent of high duty-factor electron accelerators and laser 
backscattering techniques, new Compton data have been obtained in the 
90's~\cite{Fed91} and more experiments are expected in the near 
future. The presently most accurate values for the proton 
polarizabilities were derived from the work of MacGibbon et 
al.~\cite{Mac95} whose experiments were performed with tagged photons 
at 70~MeV$\leq \nu \leq 100$~MeV and untagged ones at the higher 
energies, and analyzed in collaboration with L'vov~\cite{Lvo97} by 
means of dispersion relations (in the following denoted by DR) at 
constant~$t$. The results were 
\begin{eqnarray} 
\alpha \;&=&\; \left(12.1 \,\pm\,0.8 \,\pm\, 0.5 \right)\, 
\times\,10^{-4}\,{\mathrm fm}^3 \;, \nonumber\\ 
\beta \;&=&\; \left(2.1 \,\mp\,0.8 \,\mp\,0.5 \right)\, 
\times\,10^{-4}\,{\mathrm fm}^3 \;. 
\label{CS24} 
\end{eqnarray} 
The physics of the $\Delta$(1232) and higher resonances has been the 
objective of further recent investigations with tagged photons at 
Mainz~\cite{Mol96} and with laser-backscattered photons at 
Brookhaven~\cite{Ton98}. Such data were used to give a first 
prediction for the so-called backward spin polarizability of the 
proton~\cite{Ton98}, i.e. the particular combination 
$\gamma_{\pi}=\gamma_1+\gamma_2+2\gamma_4$ entering the Compton 
spin-flip amplitude at $\theta=180^{\circ}$, 
\begin{equation} 
\gamma_{\pi} \;=\; -\,\left[ 27.1 \,\pm\, 2.2 ({\mathrm stat + syst})\, 
{+2.8 \atop -2.4} ({\mathrm model})\right] \, 
\times\,10^{-4}\,{\mathrm fm}^4 \;. 
\end{equation} 

In 1991 Bernard et al.~\cite{Ber91} evaluated the one-loop
contributions to the polarizabilities in the framework of relativistic
chiral perturbation theory (ChPT), with the result $\alpha=10 \cdot
\beta=12.1$ (here and in the following, the scalar polarizabilities
are given in units of $10^{-4}$~fm$^3$ and the spin polarizabilities
in units of $10^{-4}$~fm$^4$).  In order to have a systematic chiral
power counting, the calculation was then repeated in heavy baryon
ChPT, the expansion parameter being an external momentum or the quark
mass. To $O(p^4)$ the result is $\alpha=10.5\pm 2.0$ and $\beta=3.5\pm
3.6$, the errors being due to 4 counter terms, which were estimated by
resonance saturation~\cite{Ber93}. One of these counter terms
describes the paramagnetic contribution of the $\Delta$(1232), which
is partly cancelled by large diamagnetic contributions of pion-nucleon
loops.  In view of the importance of the $\Delta$ resonance, Hemmert
et al.  proposed to include the $\Delta$ as a dynamical degree of
freedom.  This added a further expansion parameter, the difference of
the $\Delta$ and nucleon masses (``$\epsilon$ expansion''). A
calculation to $O(\epsilon^3)$ yielded $\alpha$ = 12.2 + 0 + 4.2 =
16.4 and $\beta$ = 1.2 + 7.2 + 0.7 = 9.1, the 3 separate terms
referring to contributions of pion-nucleon loops (identical to the
predictions of the $O(p^3)$ calculation), $\Delta$-pole terms, and
pion-$\Delta$ loops ~\cite{Hem98a,Hem97}.  These $O(\epsilon^3)$
predictions are clearly at variance with the data, in particular
$\alpha+\beta=25.5$ is nearly twice the rather precise value
determined from DR (see below).
 
The spin polarizabilities have been calculated in both relativistic
one-loop ChPT~\cite{Ber95} and heavy baryon ChPT~\cite{Hem98a}. In the
latter approach the predictions are $\gamma_0 = 4.6-2.4-0.2+0=+2.0,$
(forward spin polarizability) and $\gamma_{\pi} =
4.6+2.4-0.2-43.5=-36.7$ (backward spin polarizability), the 4 separate
contributions referring to N$\pi$-loops, $\Delta$-poles,
$\Delta\pi$-loops, and the triangle anomaly, in that order. It is
obvious that the anomaly or $\pi^0$-pole gives by far the most
important contribution to $\gamma_{\pi}$, and that it would require
surprisingly large higher order contributions to increase
$\gamma_{\pi}$ to the value of Ref.~\cite{Ton98}.  Similar conclusions
were reached in the framework of DR. Using DR at $t$ = const,
Ref.~\cite{Dre98} obtained a value of $\gamma_{\pi}=-34.3$, while
L'vov and Nathan~\cite{Lvo99} worked in the framework of backward DR
and predicted $\gamma_{\pi}=-39.5\pm2.4$.
 
As we have stated before, the most quantitative analysis of the 
experimental data has been provided by DR. In this way it has been 
possible to reconstruct the forward non spin-flip amplitude directly 
from the total photoabsorption cross section by Baldin's sum 
rule, which yields a rather precise value for the sum of 
the scalar polarizabilities 
\begin{eqnarray} 
\alpha+\beta & = & 14.2\ \ \pm\, 0.5\ \ \ {\rm(Ref.~^{54})} 
 \nonumber\\ 
  & = & 13.69 \pm\, 0.14\ \ {\rm(Ref.~^{55})}\ . 
\end{eqnarray} 
Similarly, the forward spin polarizability can be evaluated by an 
integral over the difference of the absorption cross sections in 
states with helicity 3/2 and 1/2, 
\begin{eqnarray} 
  \gamma_0 = \gamma_1-\gamma_2-2\gamma_4 & = & -1.34 \ \ 
{\rm(Ref.~^{56})}
  \nonumber \\ 
   & = & -0.6\ \ \ {\rm(Ref.~^{52})}\ . 
\end{eqnarray} 
%

\begin{figure}[htbp] 
  \centerline{\epsfxsize=10cm \epsfbox{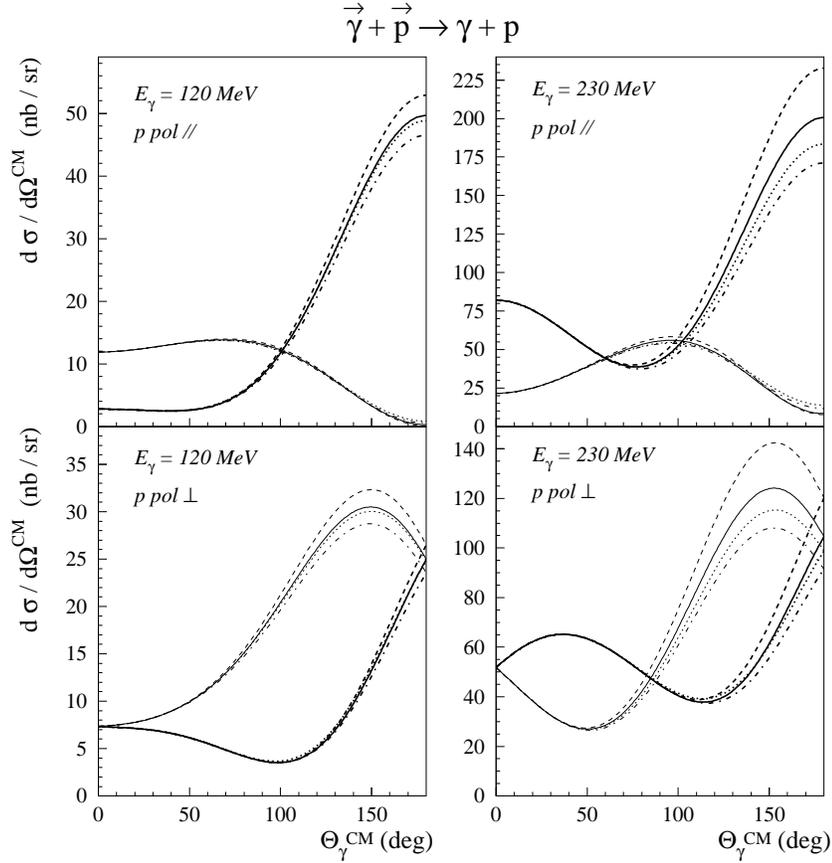}}
\vspace{-1.0cm} 
\caption 
{Double polarization cross sections for Compton scattering off the
  proton, with circularly polarized photon and target proton polarized
  along the photon direction (upper panels) or perpendicular to the
  photon direction and in the scattering plane (lower panels). The
  thick (thin) lines correspond to a proton polarization along the
  positive (negative) direction, respectively. The results of the
  dispersion calculation are for $\alpha-\beta=10$ and different
  values for $\gamma_{\pi}:\ \gamma_{\pi}=-32$ (full lines),
  $\gamma_{\pi}=-27$ (dashed lines), and $\gamma_{\pi}=-37$
  (dashed-dotted lines). The dotted line is the result for
  $\alpha-\beta=8$ and $\gamma_{\pi}=-37$. See
  Ref.~\protect\cite{Dre99c} for further details.  }
\label{fig:doublepol} 
\end{figure}

The difference can be traced back to the s-wave threshold amplitude 
$E_{0^+}(\gamma p\rightarrow n\pi^+)$, which used to be $24.9\cdot 
10^{-3}/m_{\pi}$ for the SAID~\cite{Arn96} and is $28.3\cdot 
10^{-3}/m_{\pi}$ for the HDT~\cite{Han98} multipoles, the latter value 
agreeing well with the prediction of ChPT, $28.4\cdot 
10^{-3}/m_{\pi}$~\cite{Ber92}. While these predictions relied on pion 
photoproduction multipoles, the helicity cross sections have now been 
directly determined by scattering photons with circular polarizations 
on polarized protons~\cite{Are98}. 
 
In view of the somewhat inconclusive situation, we are waiting for the
new MAMI data for Compton scattering on the proton in and above the
$\Delta$-resonance region and over a wide angular range that have been
reported preliminarily~\cite{Ahr99}.  These new data will be most
valuable to check the consistency of pion photoproduction and previous
Compton scattering results obtained at LEGS, MAMI and other
facilities.

Finally, in Fig.~\ref{fig:doublepol} we show the potential of double-
polarization observables for measuring the spin
polarizabilities~\cite{Dre99c}.  In particular, an experiment with a
circularly polarized photon and a polarized proton target should be
quite sensitive to the backward spin polarizability $\gamma_\pi$,
especially at energies between pion threshold and the $\Delta$
resonance.  In addition, possible normalization problems can be
avoided by measuring appropriate asymmetries.  Therefore such
polarization experiments hold the promise to disentangle scalar and
vector polarizabilities of the nucleon and to quantify the nucleon
spin response in an external electromagnetic field.

 
\section{PION PHOTOPRODUCTION} 
 
The reaction 
\begin{equation} 
\gamma^{\ast}(q) + N (p_1) \rightarrow \pi(p) + N(p_2) 
\end{equation} 
is described by a transition matrix element
$\varepsilon^{\mu}J_{\mu}$, with $\varepsilon^{\mu}$ the polarization
of the (virtual) photon and $J_{\mu}$ a transition current. This
current can be expressed by 6 different Lorentz structures constructed
from the independent momenta $p$, $q$ and $P=(p_1+p_2)/2$ and
appropriate Dirac matrices. Since the photon couples to the vector
current and the pion is pseudoscalar, this transition current has the
structure of an axial vector. Written in the $cm$ frame, its spacelike
($\vec{J}$) and the timelike ($\rho$) components take the form
\begin{eqnarray} 
  \vec{J} & = & \tilde{\sigma} F_1+i(\hat{q}\times\vec{\sigma}) 
  (\vec{\sigma}\cdot\hat{p}) F_2+\tilde{p} (\vec{\sigma}\cdot\hat{q}) 
  F_3 \nonumber \\ && +\tilde{p} (\vec{\sigma}\cdot\hat{p}) 
  F_4+\hat{q} (\vec{\sigma}\cdot\hat{q}) F_5+\hat{q} 
  (\vec{\sigma}\cdot\hat{p}) F_6\ , \\ \rho & = & 
  (\vec{\sigma}\cdot\hat{p}) F_7+(\vec{\sigma}\cdot\hat{q}) F_8 \ , 
\end{eqnarray} 
where $F_1$ to $F_8$ are the CGLN amplitudes~\cite{Che57}. The 
structures in front of $F_1$ to $F_6$ and $F_7$ to $F_8$ are the axial 
vectors and pseudoscalars that can be constructed from the 
$\vec{\sigma}$ matrix and the independent $cm$ momenta $\vec{p}$ and 
$\vec{q}$. We note that $\tilde{\sigma}$ and $\tilde{p}$ are the 
transverse components of $\vec{\sigma}$ and $\hat{p}$, respectively, 
with regard to $\hat{q}$.  With these definitions $F_1$ to $F_4$ 
describe the transverse, $F_5$ to $F_6$ the longitudinal and $F_7$ to 
$F_8$ the timelike components of the current.  The latter ones are 
related by current conservation, $\vec{q}\cdot\vec{J}-\omega\rho=0$, 
leading to $\mid\vec{q}\mid F_5=\omega F_8$ and 
$\mid\vec{q}\mid F_6=\omega F_7$. 
 
The CGLN amplitudes can be decomposed into a series of 
multipoles~\cite{Dre92}, 
\begin{equation} 
\{ {\cal M}_{l\pm} \} = \{ E_{l\pm},\ M_{l\pm},\ L_{l\pm} \}\ , 
\label{PP3} 
\end{equation} 
where $E$ and $M$ denote the transverse electric and magnetic 
multipoles, and $L$ are the longitudinal ones related to scalar 
(timelike, Coulomb) multipoles $S$ by current conservation. These 
multipoles are complex functions of 2 variables, e.g. ${\cal M}={\cal 
  M}(Q^2,W)$. 
 
\begin{figure}[htbp] 
\centerline{\epsfxsize=12cm \epsfbox{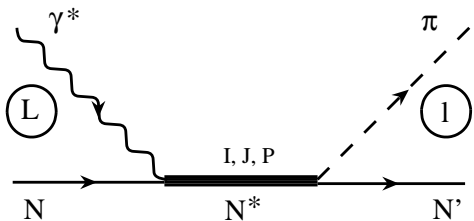}} 
\vspace{-6.0cm} 
\caption 
{Multipole notation for pion photoproduction. See text.} 
\label{figPP9} 
\end{figure}

The notation of the multipoles is clarified by Fig.~\ref{figPP9}. The
incoming photon carries the multipoles $EL$, $ML$ and $SL$, which are
contructed from its spin 1 and the orbital angular momentum.  The
parity of these multipoles is ${\cal P}=(-1)^L$ for $E$ and $S$, and
${\cal P}=(-1)^{L+1}$ for $M$. The photon is now coupled to the
nucleon with spin $1/2$ and ${\cal P}=+1$, which leads to intermediate
states with spin $J=\mid L\pm\frac{1}{2}\mid$ and the parity of the
incoming photon. The outgoing pion has negative intrinsic parity and
orbital angular momentum $l$, from which we can reconstruct the spin
$J=\mid l\pm\frac{1}{2}\mid$ and parity ${\cal P}=(-1)^{l+1}$ of the
intermediate state. This explains the notation of the multipoles,
Eq.~(\ref{PP3}), by the symbols $E$, $M$ and $S$ referring to the type
of the photon, and by the index $l\pm$ with $l$ standing for the pion
momentum and the $\pm$ sign for the two possibilities to construct the
total spin $J=\mid l\pm\frac{1}{2}\mid$ in the intermediate states.
This notation completely defines the transition, in particular it
determines the electromagnetic multipoles and the quantum
numbers of the intermediate states.
 
Let us consider as an example the excitation of the $\Delta$(1232)
with the spectroscopic notation $P_{33}$. This intermediate state
contains a pion in a $p$ wave, i.e. $l=1$ and ${\cal P}=+1$. The
indices ``33'' refer to isospin $I=3/2$ and spin $J=3/2$ respectively.
The N$\Delta$ transition can therefore take place by $M1$ or $E2$
photons, for virtual photons also $S2$ is allowed. The phase
$\delta_{l\pm}^I$ of the pion-nucleon final state is
$\delta_{1+}^{3/2}$, and the photoproduction multipoles are denoted by
$E_{1+}^{3/2}$, $M_{1+}^{3/2}$ and $L_{1+}^{3/2}$ (or $S_{1+}^{3/2}$),
i.e. in the same way as the pion-nucleon phase. As a further example,
the threshold production is determined by s-wave pions, i.e. $l=0$,
$J=\frac{1}{2}$, which leads to $E1$ or $S1$ transitions and
multipoles $E_{0+}$ or $S_{0+}$.
 
We complete the formalism of pion photoproduction by a discussion of 
isospin. Since the incoming photon has both isoscalar and isovector 
components and the produced pion is isovector, the matrix elements 
take the form~\cite{Dre92} 
\begin{equation} 
  {\cal M}_{l}^{\alpha} = \frac{1}{2}[\tau_{\alpha},\tau_0]{\cal
    M}_{l}^{(-)} + \frac{1}{2}\{\tau_{\alpha},\tau_0\}{\cal
    M}_{l}^{(+)}+\tau_{\alpha} {\cal M}_{l}^{0}\ .
\end{equation} 
The first two amplitudes on the $rhs$ can also be combined to 
\begin{equation} 
{\cal M}_{l}^{(\frac{3}{2})} = {\cal M}_{l}^{(+)} 
- {\cal M}_{l}^{(-)},\ \ \  {\cal M}_{l}^{(\frac{1}{2})} 
 = {\cal M}_{l}^{(+)} + 2{\cal M}_{l}^{(-)}\ , 
\end{equation} 
where the upper index $\frac{3}{2}$ or $\frac{1}{2}$ denotes the
isospin of the final state. The 4 physical amplitudes are then given
in terms of linear combinations of the 3 isospin amplitudes. We note,
however, that the isospin symmetry is broken by the mass differences
between the nucleons $(n, p)$ and pions $(\pi^\pm, \pi^0)$ and by
explicit Coulomb effects, in particular near threshold.

\subsection{Threshold pion photoproduction} 
As has been pointed out before, threshold production is dominated by 
the multipoles $E_{0+}$ ($s$-wave pions). For these multipoles there 
existed a venerable low energy theorem~\cite{deB70}, which however had 
to be revised in view of surprising experimental evidence. 
 
\begin{table}[t] 
\begin{center} 
\caption{The s-wave amplitude $E_{0+}$ at threshold in units of 
   $10^{-3}/m_{\pi}$. See text.} 
\label{tab:e0p_thr} 
\vspace{0.5cm} 
\begin{tabular}{|c|c|c|c|} 
\hline 
         & $\gamma p \rightarrow \pi^+n$ & $\gamma n\rightarrow \pi^-p$ 
         & $\gamma p \rightarrow \pi^{0}p$ \\ \hline 
``LET''~\cite{deB70}  & 27.5                          & -32.0  & -2.4 \\ 
ChPT~\cite{Ber96} & $28.2 \pm .6$                 & $-32.7 \pm .6$  & -1.16 \\ 
DR~\cite{Han97} & 28.4                          & -31.9  & -1.22 \\ 
experiment & $28.3 \pm .2$~\cite{Ada76} 
 & $-31.8 \pm .2$~\cite{Ada76} & -1.31$\pm$.08~\cite{Fuc96,Berg96} \\ 
\hline 
\end{tabular} 
\end{center} 
\end{table} 
Table~\ref{tab:e0p_thr} compares our predictions from dispersion
theory to the ``classical'' low energy theorem (LET), ChPT and
experiment. Note that ChPT~\cite{Ber96} contains the lowest order loop
corrections, while ``LET'' is based on tree graphs only.  Due to the
coupling between the channels, the real part of $E_{0+}(\gamma
p\rightarrow \pi^0p)$ obtains large contributions from the imaginary
parts of the higher multipoles via the dispersion integrals.
Altogether these contributions nearly cancel the large contribution of
the Born terms, which correspond to the result of pseudoscalar
coupling, leading to a total threshold value~\cite{Han97}
\begin{eqnarray} 
\mbox{Re}E_{0+}^{\mbox{\scriptsize thr}}(p\pi^0) 
& = & -7.63 + 4.15 - 0.41 + 2.32 + 0.29 + 0.07 = -1.22, \nonumber \\ 
\mbox{Re}E_{0+}^{\mbox{\scriptsize thr}}(n\pi^{0}) 
& = & -5.23 + 4.15 - 0.41 + 3.68 - 0.93 - 0.05 = \ 1.19, 
\end{eqnarray} 
where the individual contributions on the $rhs$ are, in that order, 
the Born term, $M_{1+}, E_{1+}, E_{0+}, M_{1-}$ and higher multipoles. 
 
As we see from Table~\ref{tab:e0p_thr}, the discrepancy between the 
``classical'' LET and the experiment is very substantial in the case 
of $\pi^0$ production on the proton. The reason for this was first 
explained in the framework of ChPT by pion-loop corrections. An 
expansion in the mass ratio $\mu=m_{\pi}/M\approx1/7$ leads to the 
result~\cite{Ber91b} 
\begin{equation} 
E_{0+} (\pi^0 p) = \frac{eg_{\pi N}}{8\pi m_{\pi}} 
\left \{ \mu-\mu^2 \frac{3+\kappa_p}{2}- 
\mu^2 \frac{M^2}{16f^2_{\pi}} +\ ...\right \}\ , 
\label{PP6} 
\end{equation} 
where $g_{\pi N}$ is the pion-nucleon coupling constant and
$f_{\pi}\approx93$~MeV the pion decay constant. We observe that the
leading term is proportional to $\mu$, which suppresses this process
relative to charged pion production. The leading terms of these
expansions can be understood, to some degree, by simply relating the
dipole moments in the respective pion-nucleon states. In particular
the expansion for $\gamma n\rightarrow\pi^0 n$ starts at $O(\mu^2)$,
because both particles in the final state are neutral. The third term
on the $rhs$ of Eq.~(\ref{PP6}) is the loop correction.  Though
formally of higher order in $\mu$, its numerical value is larger than
the leading term!
 
While the threshold cross section receives its forward-backward
asymmetry essentially from the combination
${\mbox{Re}}\{E_{0^+}^{\ast}(M_{1^+}+ 3E_{1^+}-M_{1^-})\}$, the photon
asymmetry $\Sigma$ is dominated by
${\mbox{Re}}\{M_{1^+}^{\ast}(E_{1^+}+M_{1^-})\}$ and the target
asymmetry $T$ by ${\mbox{Im}}\{E_{0^+}^{\ast} (E_{1^+}-M_{1^+})\}$.
Since $E_{1^+}$ is small, the value of $\Sigma$ is surprisingly
sensitive to the multipole $M_{1^-}$ resonating at the Roper resonance
$\mbox{N}^{\ast}$(1440). The observable T, on the other side, measures
the phase of pion-nucleon s-wave scattering at threshold relative to
the phase of the $\Delta$ (1232) multipole.

\begin{figure}[ht] 
\centerline{\epsfxsize=10cm \epsfbox{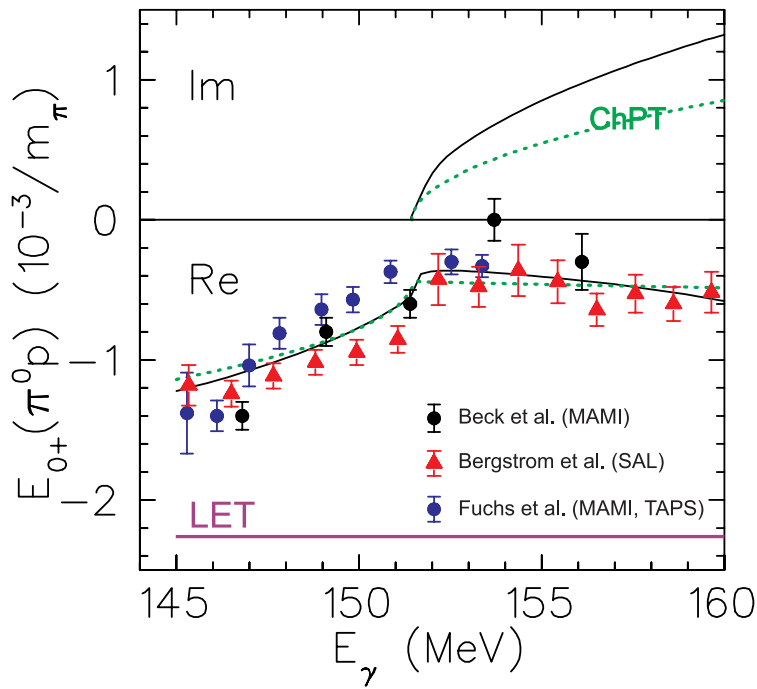}} 
\vspace{-3.0cm} 
\caption 
{The real (Re) and imaginary (Im) parts of the threshold amplitude
  $E_{0+}(p\pi^0)$ as predicted by dispersion
  relations~\protect\cite{Han97} (full lines), chiral perturbation
  theory~\protect\cite{Ber96} (dashed lines), and the ``classical''
  low energy theorem~\protect\cite{deB70} (LET). Experimental data
  from MAMI~\protect\cite{Fuc96,Bec90} and SAL~\protect\cite{Ber96}.}
 \label{figPP10} 
\end{figure}

Finally, the energy dependence of $E_{0+}(\pi^0p)$ near threshold is 
shown in Fig.~\ref{figPP10}. The discrepancy between the ``classical'' 
LET and the experimental data is clearly seen, and one also observes a 
``Wigner cusp'' at the $\gamma p\rightarrow\pi^+n$ threshold. In 
particular, the imaginary amplitude rises sharply due to the strong 
coupling to this channel. Since charged pion production is much more 
likely to happen, neutral pions will often be produced by rescattering 
$\gamma p\rightarrow \pi^+n\rightarrow\pi^0p$. 
 
\subsection{Pion production in the resonance region} 
 
The search for a deformation of the ``elementary'' particles is a 
longstanding issue. Such a deformation is evidence for a strong tensor 
force between the constituents, originating in the case of the nucleon 
from the residual force of gluon exchange. Depending on one's 
favourite model, such effects can be described by d-state admixture in 
the quark wave function~\cite{Isg78}, tensor correlations between the 
pion cloud and the quark bag~\cite{The80,Kae83}, or by exchange 
currents accompanying the exchange of mesons between the 
quarks~\cite{Buc97}. Unfortunately, it would require a target with a 
spin of at least 3/2 (e.g. $\Delta$ matter) to observe a static 
deformation.  An alternative is to measure the transition quadrupole 
moment between the nucleon and the $\Delta$, i.e.  the amplitude 
$E_{1+}$, which is sensitive to model parameters responsible for a 
possible deformation of the hadrons. 
 
The experimental quantity of interest is the ratio $R_{EM} = 
E_{1+}/M_{1+}$ in the region of the $\Delta$. The two amplitudes 
$E_{1+}$ and $M_{1+}$ are related to the helicity amplitudes, which may 
be determined by scattering an incident photon with circular 
polarization off a target nucleon with its spin oriented in the 
direction or opposite to the photon momentum $\vec{q}$, 
\begin{eqnarray} 
  A_1 &=& A_{1/2} = \frac{1}{\sqrt{2q}} \left\langle N^{\ast} (J, M = 
    \frac{1}{2}) \mid J_+ \mid N (J_i = \frac{1}{2}, M_i = - 
     \frac{1}{2}) \right\rangle \nonumber \\  A_3 
  &=& A_{3/2} = \frac{1}{\sqrt{2q}}\left\langle N^{\ast} 
  (J, M = \frac{3}{2}) \mid J_+ 
    \mid N (J_i = \frac{1}{2}, M_i = + \frac{1}{2}) \right\rangle, 
\end{eqnarray} 
where $J_+$ is the hadronic current corresponding to the absorption of 
a photon with positive helicity on the nucleon $N$ with spin $J_i = 
\frac{1}{2}$ and spin projection $M_i$, leading to a resonance state 
$N^{\ast}$ with spin $J \ge \frac{1}{2}$ and spin projection $M$.  It 
is obvious that all resonances can generally contribute to $A_1$, 
while only resonances with $J \ge \frac{3}{2}$ will contribute to 
$A_3$. The helicity-conserving process $A_1$ can also occur on an 
individual (massless) quark, whereas $A_3$ is forbidden in that 
approximation.  Hence perturbative QCD predicts that $A_3$ should 
vanish for high momentum transfer, i.e. for electroproduction and 
$Q^2=\vec{q}^{\ 2}-\omega^2 \rightarrow \infty$. 
 
We shall now compare the prediction of K\"albermann and Eisenberg~ 
\cite{Kae83} with our analysis of the modern pion photoproduction 
data~\cite{Han98,Han97}.  The helicity amplitudes for the $N 
\rightarrow \Delta$ transition will be given in the usual units of 
10$^{-3}$~GeV$^{-\frac{1}{2}}$, and the prediction of 
Ref.~\cite{Kae83} is obtained for a bag radius of 1~fm, which was the 
preferred value in the 1980's.  The result is 
\begin{equation} 
\begin{array}{ll} 
{\rm {Ref.~^{74}:}} \ \
 A_1 = -130  & A_3 = - 250 \nonumber \\ 
{\rm {Ref.~^{58}:}} \ 
A_1 = -131 \pm 1\ \  & A_3 = - 252\pm 1\ , 
\label{PP8} 
\end{array} 
\end{equation} 
the agreement being truly astounding though somewhat accidental, 
because the theoretical value depends on the bag radius. However, the 
result is relatively stable, even a drastic decrease of the bag radius 
to 0.6 fm will change the helicity amplitudes by only 10\%. This 
success of the chiral bag model is even more outstanding when 
compared with the results of the quark model without pionic degress 
of freedom. From a selection of ten quark model calculations published 
over the past 20 years we find $- 113 \le A_1 \le -82$ and $- 195 
\le A_3 \le -58$, values far off the experimental data. 
 
The helicity amplitudes are related to the electric and magnetic 
multipoles, 
\begin{eqnarray} 
  M_{1^+} & = & -\frac{1}{2\sqrt{3}}(\sqrt{3}\ A_{1/2} + 3\ A_{3/2})\ 
  , \nonumber \\ E_{1^+} & = & -\frac{1}{2\sqrt{3}} (\sqrt{3}\ 
  A_{1/2}-A_{3/2})\ .
\label{PP9} 
\end{eqnarray} 
Since $A_{\frac{3}{2}} \simeq \sqrt{3}\ A_{\frac{1}{2}}$ according to 
Eq.~(\ref{PP8}), the model predicts that $E_{1^+}$ (electric 
quadrupole excitation $E2$) is very much smaller than $M_{1^+}$ 
(magnetic dipole excitation $M1$).  A few years after the pioneering 
work of Ref.~\cite{Kae83}, we obtained, for a bag radius of 0.6 fm, 
the ratio $R = E_{1^+}/M_{1^+} = - 2.8 \%$ \cite{Ber88}. This result 
differed by a factor of two from the then accepted experimental value 
$R_{1988} \simeq - 1.3 \%$.  However, it is quite close to the recent 
MAMI data of Beck et al.~ \cite{Bec97}, $R_{1997}= (- 2.5 \pm 0.2 \pm 
0.2) \%$, and to our global analysis of the data \cite{Han98}, 
$R_{1998} = (-2.5 \pm 0.1) \%$. 

\begin{figure}[htbp] 
\centerline{\epsfxsize=10cm \epsfbox{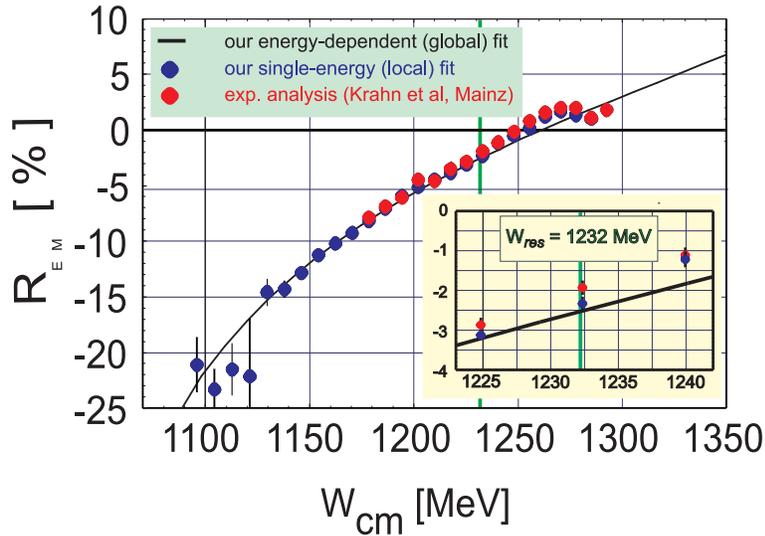}} 
\caption 
{The ratio $R_{EM}=R$ as function of the energy $W_{cm}$ of the
  pion-nucleon system. See Refs.~\protect\cite{Han98,Bec97} and
  references given therein }
\label{figPP11} 
\end{figure}

As may be seen from Fig.~\ref{figPP11} , the ratio $R = R_{EM}$
changes rapidly with the energy $W_{cm}$ of the pion-nucleon system.
The reason for this energy dependence is the nonresonant background,
which is particularly large in the case of the small $E_{1^+}$
multipole. The historically first prediction of that number is due to
Chew et al.~\cite{Che57} in 1957 who found $R \simeq 0$ from a
dispersion theoretical analysis. Such value was later explained by
Becchi and Morpurgo~\cite{Bec65} in the framework of the constituent
quark model.  In the following years the quark models were refined by
introducing tensor correlations, with the result of finite, small and
usually negative values for $R$. Such correlations have been motivated
in different ways, by hyperfine interactions between the quarks
\cite{Isg78}, pion-loop effects \cite{Kae83} and, more recently,
exchange currents \cite{Buc97}. In analogy with heavy even-even nuclei
having ``intrinsic'' deformation, finite values of $E2$ are often
referred to as "bag deformation" or "deformation of the nucleon",
although a quadrupole moment cannot be observed for an object with
spin $J < 1$. Ideally one could probe the static quadrupole moment of
the $\Delta$ by experiments like $\pi N \rightarrow \Delta \rightarrow
\Delta \gamma \rightarrow \pi N \gamma$, however a closer look shows
that this is hardly a realistic possibility. In conclusion it is
precisely the $N \Delta$ transition quadrupole moment that provides us
with information on tensor correlations in the nucleon, which can be
translated, e.g., into a $d$-state admixture in the quark wave
function.  In such a model the $\Delta$ would have an oblate
deformation, much smaller than a frisbee and much larger than the
earth, in absolute numbers quite comparable to the deuteron, which
however has a prolate deformation.

\begin{figure}[htbp] 
\centerline{\epsfxsize=9cm \epsfbox{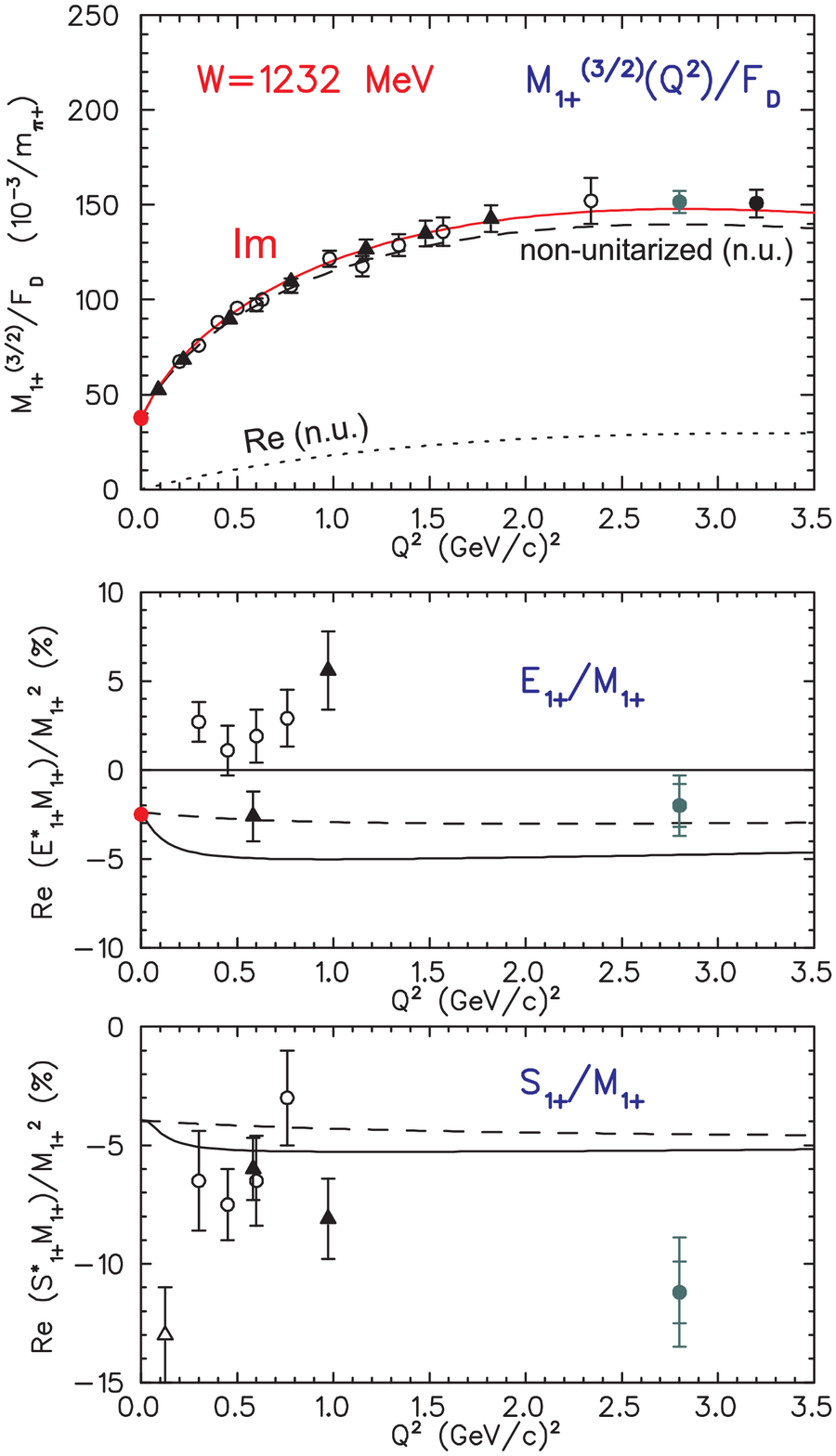}} 
\caption 
{The amplitudes for $\Delta$ excitation as function of $Q^2$. Top:
  $M_{1+}$ divided by the dipole form factor $F_D=G_D$, center: the
  ratio $E_{1+}/M_{1+}$, bottom: the ratio $S_{1+}/M_{1+}$. See text
  and Ref.~\protect\cite{Dre99a}. }
\label{figPP12} 
\end{figure}

Meson electroproduction allows us to study the dependence of the
multipoles on momentum transfer, $M_{l\pm} = M_{l \pm} (Q^2)$, i.e. to
probe the spatial distribution of the transition strength.  In
addition, the virtual photon carries a longitudinal field introducing
a further multipole, $S_{l\pm}$. The $Q^2$ dependence of the $N\Delta$
multipoles is displayed in Fig.~\ref{figPP12}. In the top figure, we
show the results for the magnetic multipole divided by the standard
dipole form factor. The data are compared to the predictions of
our unitary isobar model (UIM)~\cite{Dre99a}. This model contains the
usual Born terms, vector meson exchange in the t-channel and nucleon
resonances in the s-channel, unitarized partial wave by partial wave
with the appropriate pion-nucleon phases and inelasticities. The
center piece of Fig.~\ref{figPP12} shows the ratio $R = R (Q^2)$
compared to mostly older and strongly fluctuating data.  More recent
data from Jefferson Lab~\cite{Sto98} indicate, however, that even at
$Q^2 \approx 2.8$ and 4 (GeV/c)$^2$ this ratio remains negative and of
the order of a few per cent. This is surprising, because perturbative
QCD predicts that the helicity amplitude $A_3$ should vanish for $Q^2
\rightarrow \infty$ and, hence, the ratio $R$ should approach +100\%
(see Eq.~(\ref{PP9})). Finally, the bottom figure shows the
corresponding longitudinal-transverse ratio $S_{1^+}/M_{1^+}$.  Recent
experimental data at ELSA, MAMI and MIT~\cite{Got98} at $Q^2 \approx
0.5$ (GeV/c$)^2$ yield ratios of about -7 \%, slightly below our
prediction, while the preliminary data from the Jefferson
Lab~\cite{Sto98} at larger $Q^2$ seem to indicate considerably lower
values between -10~\% and -20~\%. From perturbative QCD one expects
that this ratio should vanish for $Q^2 \rightarrow \infty$.
 
The modern precision experiments will be continued to the higher 
resonances. Concerning the $N^{\ast}$(1440) or Roper resonance, both 
data and predictions are still in a deploratory state, and it will 
require double-polarization experiments to find out about the nature 
of that resonance. One possibility to tackle the problem will be pion 
production by linearly polarized photons on longitudinally polarized 
protons.  Such an experiment measures the polarization observable $G 
\sim Im M_{1^-} Re M_{1^+}$, i.e. an interference of the $\Delta$ 
resonance with the absorptive part of the Roper multipole $M_{1^-}$.

\begin{figure}[htbp] 
\centerline{\epsfxsize=6.5cm \epsfbox{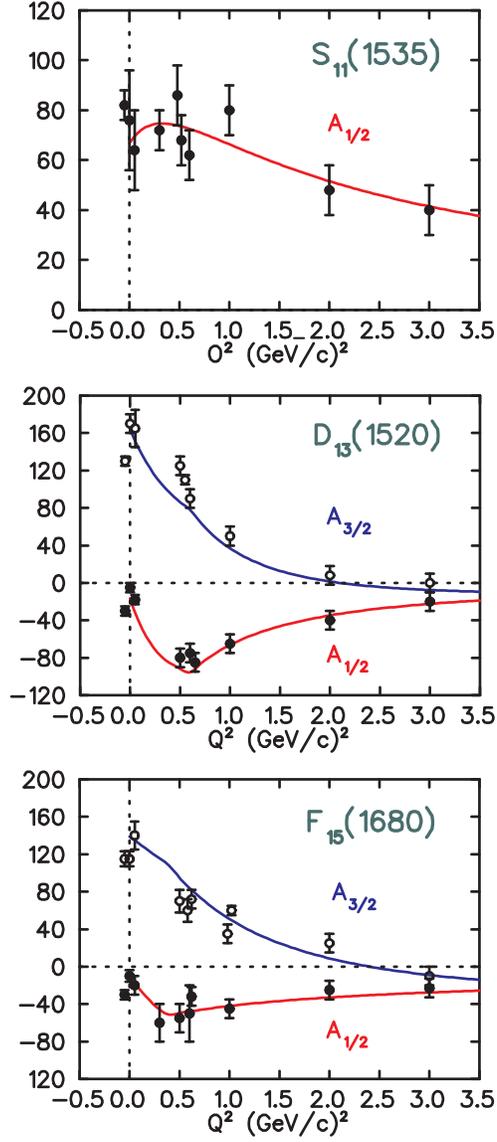}} 
\caption 
{The helicity amplitudes $A_{1/2}^p$ and $A_{3/2}^p$ for the 
  resonances $S_{11}$(1535), $D_{13}(1520)$, and $F_{15}(1680)$ as 
  functions of $Q^2$. See text and Ref.~\protect\cite{Dre99a}.} 
\label{figPP13} 
\end{figure}

The existing information on some of the higher resonances is shown in 
Fig.~\ref{figPP13}. For our discussion in the next chapter it is 
important to note: 
\begin{enumerate} 
\item [(i)] Because of its quantum numbers $J^P = \frac{1}{2}^-$, the 
  resonance $S_{11}$(1535) is only excited via the $A_{1/2}$ 
  amplitude. As function of $Q^2$, this amplitude drops much slower 
  than any other resonance of the nucleon. With a resonance position 
  very close to $\eta$ production threshold, the $S_{11}$(1535) has 
  an $\eta$ branching ratio of about 50~\%, while this ratio is of 
  the order of 1~\% or less for all other resonances. 
\item[(ii)] The resonances $D_{13}$(1520) and $F_{15}$(1680) carry
  most of the electric dipole and quadrupole strengths, respectively.
  For real photons $(Q^2 = 0)$ their helicity amplitudes $A_{1/2}^p$
  are nearly zero, but already at $Q^2 \approx 0.5$ (GeV/c)$^2$
  $A_{1/2}^p$ and $A_{3/2}^p$ are of equal importance, and in
  accordance with pQCD, $A_{3/2}^p$ decreases rapidly for $Q^2
  \rightarrow \infty$.
\end{enumerate}

 
\section{SUM RULES} 
 
As has been stated in Eq.~(\ref{CS21}), the GDH sum rule 
connects the integral 
\begin{equation} 
I = \int_{\nu_0}^{\infty} \frac{\sigma_{1/2} (\nu) - \sigma_{3/2} 
  (\nu)}{\nu} d\nu 
\end{equation} 
with the anomalous magnetic moment. 
 
On the basis of the pion-nucleon multipoles and certain assumptions 
for the higher channels, various authors have estimated this integral. 
As shown in Table~\ref{tabSR.1}, the absolute value of the proton 
integral $I_p$ has been consistently overpredicted, while the neutron 
integral $I_n$ comes out too small.  This has the consequence that not 
even the sign of the isovector combination $I_p-I_n$ agrees with the 
sum rule value.  This apparent discrepancy has led to speculations 
that the GDH integral should not converge for various reasons, e.g. 
due to a generalized current algebra, because of fixed axial vector 
poles or influences of the Higgs particle. None of these arguments is 
too convincing at present. In fact one should realize that the GDH 
integrand is an oscillating function of photon energy, with multipole 
contributions of alternating sign. Therefore, little details matter 
and a stable result requires very exact data. Comparing again the 
results obtained with the SAID~\cite{Arn96} and HDT~\cite{Han98} 
multipoles, the generally accepted threshold value of $E_{0^+}$ 
reduces the ``discrepancy'' with the sum rule value by about $25\%$ 
(see Table~\ref{tabSR.1} and Ref.~\cite{Dre98}).

\begin{table}[htbp] 
\begin{center} 
\caption{Predictions for the GDH integral for proton $(I^p$), 
  neutron $(I^n$), and the difference $I^p-I^n$ in units of $\mu$b.
  With the exception of Ref.~\protect\cite{Bur93}, the two-pion
  contribution has been taken from Ref.~\protect\cite{Kar73}.}
\label{tabSR.1} 
\vspace{0.5cm} 
\begin{tabular}{|l|r|r|r|} 
\hline 
                 & $I^p$    & $I^n$    & $I^p-I^n$     \\ 
\hline 
GDH              & -205     & -233     & 28           \\ 
\hline 
Ref.~\cite{Kar73}     & -261     & -183     & -78          \\ 
\hline 
Ref.~\cite{Wor92}     & -260     & -157     &-103          \\ 
\hline 
Ref.~\cite{Bur93}     & - 223    &          &              \\ 
\hline 
Ref.~\cite{San94}     & - 289    & -160     & - 129         \\ 
\hline 
Ref.~\cite{Dre98}     &-261      & -180     & -81          \\ 
\hline 
\end{tabular} 
\end{center} 
\end{table}

\begin{figure}[ht] 
\centerline{\epsfxsize=7cm \epsfbox{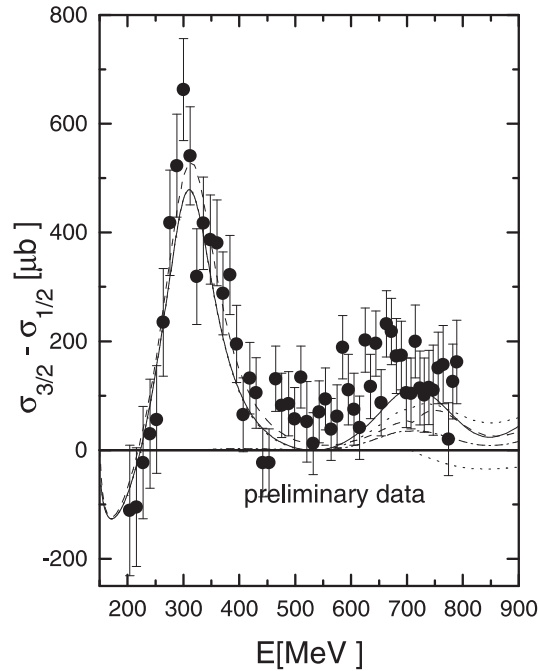}} 
\caption 
{The difference of the helicity cross section,
  $\sigma_{3/2}-\sigma_{1/2}$, as function of the $lab$ energy $\nu=E$
  of the photon. The theoretical predictions of
  Refs.~\protect\cite{Dre99a,Are97} are compared to the preliminary
  data of the 1998 MAMI experiment$^{57}$.}
\label{figSR14} 
\end{figure}

The first direct measurement of the helicity cross sections was
recently performed at MAMI in the energy region 200~MeV
$<\nu<800$~MeV~\cite{Are98}. The experiment will be extended to the
higher energies at ELSA. Some preliminary results are shown in
Fig.~\ref{figSR14}, which contains only 5~\% of the data taken in the
1998 run. The figure shows the importance of charged pion production
near threshold (multipole $E_{0+}$), and the dominance of the
multipole $M_{1+}^{3/2}$ in the $\Delta$ resonance region. At yet
higher energies the data lie above the prediction for one-pion
production, which indicates considerable two-pion contributions.
These data establish that the forward spin polarizability should be
$\gamma_0\approx -0.8\cdot 10^{-4}$~fm$^{4}$. Furthermore the
preliminary data saturate the GDH sum rule at $\nu\approx 800$~MeV if
one accepts our predictions~\cite{Dre98} for the energy range between
threshold and 200~MeV. However, more data are urgently required at
energies both below 200~MeV and above 800~MeV.

In view of the difficulty to obtain even the proper sign for the
proton-neutron difference from the older data (see
Table~\ref{tabSR.1}), it is of considerable interest to measure the
GDH for the neutron. However, such investigations are difficult due to
nuclear binding effects.  While it is generally assumed that $^2$H and
$^3$He are good neutron targets, the sum rule requires to integrate
over all regions of phase space and not only the region of quasifree
kinematics. In fact there exists a GDH sum rule for systems of any
spin, and hence every nucleus should have a well-defined value for the
GDH integral.  With the definition of Table~\ref{tabSR.1}, one finds
the small value $I(^2$H$)=-0.65~\mu$b due to the fact that the deutron
lies very close to the Schmidt line.  However, a loosely bound system
of neutron and proton would be expected to have $I_p+I_n=-438~\mu$b,
which differs by 3 orders of magnitude from the deuteron value!
Obviously the large contributions from pion production have to be
canceled by binding effects in the deuteron.  As has been shown by
Arenh\"ovel and collaborators~\cite{Are97}, such contribution is mainly
due to the transition from the $^3$S$_1$ ground state of $^2$H to the
$^1$S$_0$ resonance at 68~keV. Weighted with the inverse power of
excitation energy, the absorption cross section for this low-lying
resonance cancels the huge cross sections due to pion production. We
note that the opposite sign of the two contributions is due to the
fact that the spins of the 3 quarks become aligned by the transition
$N\rightarrow\Delta$, while the nucleon spins are parallel in the
deuteron $(^3$S$_1$) but antiparallel in the $^1$S resonance.
However, in addition to this low lying resonance, there are also
sizeable sum rule contributions by break-up reactions
$\gamma+d\rightarrow n+p$ in the range below and above pion threshold.
Such effects are usually triggered by meson exchange currents (the
virtual pions below or the real pions above threshold are reabsorbed
by the other nucleon) or isobaric currents (a $\Delta$ is produced but
decays by final state interaction without emitting a pion). In
addition there are also contributions of coherent $\pi^0$ production,
i.e.  $\gamma+d\rightarrow\pi^0+d$. It is general to all these
processes that they cannot occur on a free nucleon, though they are
certainly driven by pion production and nucleon resonances. This leads
to the serious question: Which part of the GDH integral is nuclear
structure and hence should be subtracted, and how should one divide
the rest into the contributions of protons and neutrons? The problem
is not restricted to the deuteron but quite general. For example the
``neutron target'' $^3$He has the same sum rule as the nucleon except
that one has to replace $e,\ m$ and $\kappa$ of the nucleon by the
charge $(Q=2e)$, mass $(M\approx3m_N)$ and anomalous magnetic moment
of $^3$He. The result is $I(^3$He$)=-496~\mu$b, while we naively
expect $I_n=-233~\mu$b, because the spins of the two protons are
antiparallel and hence should not contribute to the helicity
asymmetry.

These considerations can be generalized to virtual photons by electron 
scattering. While the coincidence cross section for the reaction 
$\vec{e}+\vec{p}\rightarrow e'+N+\pi$ contains 18 different response 
functions~\cite{Dre92}, only 4 responses remain after integration over 
the angles of pion emission, which is exactly the result of 
Eq.~(\ref{KIN9}).  The 4 partial cross sections can in principle be 
separated by a super-Rosenbluth plot if one varies the polarizations. 
These are the transverse polarizations $\varepsilon$ of the virtual 
photon, the polarization $P_e$ of the electron ($\pm 1$ for 
the relativistic case), and the nucleon's polarization in the 
scattering plane of the electron, with components $P_z$ in the 
direction and $P_x$ perpendicular to the virtual photon momentum. 
 
The multipole content of one-pion production to the partial 
cross sections is~\cite{Dre95,Dre99b} 
\begin{eqnarray} 
\sigma_T^{(1\pi)} & = & 4\pi\frac{|{\bf 
    k}_{\pi}^{cm}|}{k^{cm}} 
\sum_l\frac{1}{2}(l+1)^2  \\ 
&& \cdot [(l+2)(| E_{l+}|^2 + | 
M_{l+1,-}|^2) + l(| M_{l+}|^2 + | E_{l+1,-}|^2)] \nonumber 
\\ & = & 4\pi\frac{|{\bf 
    k}_{\pi}^{cm}|}{k^{cm}} \{| E_{0+}|^2 + 2| 
M_{1+}|^2 + 6| E_{1+}|^2 + | M_{1-}|^2 + 2| 
E_{2-}|^2 \pm\ ...\}\ ,\nonumber \\ 
\sigma_L^{(1\pi)} & = & 4\pi\frac{|{\bf 
    k}_{\pi}^{cm}|}{k^{cm}} \left (\frac{Q}{\omega^{cm}}\right 
  )^2  \sum_l\frac{1}{2}(l+1)^3 [|L_{1+}|^2 + |L_{l+1,-}|^2] \\ 
& = & 4\pi\frac{|{\bf 
    k}_{\pi}^{cm}|}{k^{cm}} \left (\frac{Q}{\omega^{cm}}\right 
  )^2 \left \{|L_{0+}|^2 + 8|L_{1+}|^2 + |L_{1-}|^2 + 8|L_{2-}|^2 
   \pm\ ... \right \} 
  \ , \nonumber \\ 
\sigma_{TT'}^{(1\pi)} & = & 4\pi\frac{|{\bf 
    k}_{\pi}^{cm}|}{k^{cm}} 
\sum_l\frac{1}{2}(l+1) [-(l+2)(|E_{l+}|^2 + |M_{l+1,-}|^2)  \\ 
&& + l(|M_{l+}|^2 + |E_{l+1,-}|^2) - 2l(l+2) (E_{l+}^{\ast}M_{l+}- 
     E^{\ast}_{l+1,-}M_{l+1,-})] \nonumber \\ 
& = & 4\pi\frac{|{\bf k}_{\pi}^{cm}|}{k^{cm}} 
\{-|E_{0+}|^2 + |M_{1+}|^2 - 6E_{1+}^{\ast}M_{1+} - 
3|E_{1+}|^2 + |E_{2-}|^2 \pm\ ... \}\, \nonumber \\ 
\sigma_{LT'}^{(1\pi)} & = & 4\pi\frac{|{\bf 
    k}_{\pi}^{cm}|}{k^{cm}} \left (\frac{Q}{\omega^{cm}}\right 
  )  \sum_l\frac{1}{2}(l+1)^2 \\ 
&& \cdot [-L_{l+}^{\ast}((l+2)E_{l+} + lM_{l+}) + L_{l+1,-}^{\ast} 
   (lE_{1+,-} + (l+2)M_{l+1,-})] \nonumber \\ 
& = & 4\pi\frac{|{\bf 
    k}_{\pi}^{cm}|}{k^{cm}} \left (\frac{Q}{\omega^{cm}}\right 
  )\{-L_{0+}^{\ast}E_{0+} - 2L_{l+}^{\ast}(M_{1+}+3E_{1+}) \nonumber \\ 
&&  + L_{1-}^{\ast}M_{1-} + L_{2-}^{\ast}E_{2-}\pm\ ...\}\ .  \nonumber 
\end{eqnarray} 
Since the partial wave decomposition is defined in the hadronic $cm$
frame, the appropriate $cm$ values of the kinematical observables have
to be used in these equations, in particular the $cm$ momentum and the
$cm$ energy of the virtual photon, $k^{cm}=\frac{m}{W}k$ and
$\omega^{cm} = \frac{1}{W}\sqrt{m^2\nu^2-Q^2(W^2-m^2)}$ respectively.
We note that $\omega^{cm}$ has a zero if $W=\sqrt{m^2+Q^2}$, which is
compensated by a corresponding zero in the longitudinal multipole.
This situation can be avoided by using the ``scalar'' multipoles
(rather to be called ``Coulomb'' or ``time-like'' multipoles!),
\begin{equation} 
S_{l\pm} = \frac{k^{cm}}{\omega^{cm}}L_{l\pm}\ . 
\end{equation} 

While $\sigma_T$ and $\sigma_L$ are the sum of squares of transverse 
$(E_{l\pm}, M_{l\pm})$ and longitudinal $(L_{l\pm})$ multipoles 
respectively, the interference structure functions $\sigma'_{TT} = 
(\sigma_{3/2}-\sigma_{1/2})/2$ and $\sigma'_{LT}$ contain multipole 
contributions of alternating sign. The multipoles involved are now 
functions of energy and momentum transfer, ${\cal M}_{l\pm}={\cal 
  M}_{l\pm}(\nu, Q^2)$. 
 
The 4 cross sections are related to the familiar structure functions 
of deep inelastic lepton scattering, 
 
\begin{equation} 
\{\sigma_T\ , \sigma_L\ ;\ \sigma'_{LT}\ , \sigma'_{TT}\}\Longrightarrow 
\{F_1\ , F_2\ ; G_1\ , G_2\}\ . 
\label{SR5} 
\end{equation} 
In the Bjorken scaling region, the 2 arguments $\nu$ and $Q^2$ of 
these functions can be replaced by the scaling variable $x=Q^2/2m\nu$, 
which leads to the definition of quark distribution functions. For the 
spin structure functions $G_1$ and $G_2$ we find 
 
\begin{eqnarray} 
\label{SR6} 
g_1(\nu, Q^2)& = &\frac{\nu}{m}\ G_1(\nu, Q^2)\rightarrow \ 
g_1(x) = \frac{1}{2}\sum e^2_i (f_i^{\uparrow}-f_i^{\downarrow}) \nonumber \\ 
g_2(\nu, Q^2)& = &\frac{\nu^2}{m^2}\ G_2(\nu, Q^2)\rightarrow 
g_2(x) = \frac{1}{2}\sum e^2_i (f_i^{\rightarrow}-f_i^{\leftarrow})\ , 
\end{eqnarray} 
where the arrows indicate the different directions of the quark spins. 
With these definitions we can express a set of generalized sum rules 
in terms of both the quark spin functions of Eq.~(\ref{SR6}) and the cross 
sections of Eq.~(79), e.g. 
 
\begin{eqnarray} 
\label{SR7} 
I_1(Q^2)& = &\int_{\nu_0}^{\infty}\frac{d\nu}{\nu}\ G_1(\nu,Q^2) = 
\frac{2m}{Q^2} \int_0^{x_0}dx\ g_1 (x,Q^2)\rightarrow 
\frac{2m^2}{Q^2} \Gamma \nonumber \\ 
& = & \frac{m^2}{2\pi e^2}\int^{\infty}_{\nu_0} d\nu\ (1-x)\ 
\left (\sigma_{1/2}-\sigma_{3/2}+2\frac{Q}{\nu}\sigma'_{LT} \right )\ , 
\end{eqnarray} 
 
\begin{eqnarray} 
\label{SR8} 
I_2(Q^2)& = &\frac{1}{m} \int^{\infty}_{\nu_0} d\nu\ G_2(\nu,Q^2) = 
\frac{2m^2}{Q^2}\int^{x_0}_0 dx\ g_2 (x,Q^2) \nonumber \\ 
& = & \frac{m^2}{2\pi e^2}\int^{\infty}_{\nu_0} d\nu\ (1-x)\ 
\left (-\sigma_{1/2}+\sigma_{3/2}+2\frac{\nu}{Q}\sigma'_{LT} \right )\ , 
\end{eqnarray} 
with $\nu_0$ and $x_0$ the lowest threshold for inelastic reactions. 
 
Eq.~(\ref{SR7}) is a possible generalization of the GDH sum rule, 
because $I_1(0) = -\kappa^2/4$. However, a large variety of 
generalized GDH sum rules can be obtained by adding different 
fractions of the interference term $\frac{Q}{\nu}\sigma'_{LT}$, which 
vanishes both in the real photon limit, $Q^2\rightarrow 0$, and in the 
asymptotic region, $Q^2\rightarrow\infty$. The most obvious choice 
would be to simply drop this term in Eq.~(\ref{SR7}). As it stands, 
however, the definition of $I_1$ is the natural definition of an 
integral over the spin structure function $g_1$. In particular it has 
the asymptotic behaviour indicated in Eq.~(\ref{SR7}), with 
$\Gamma$ a constant.  The fact that the experimental value of $\Gamma$ 
differed from earlier predictions~\cite{Ell74} led to the ``spin 
crisis'' and taught us that less than half of the nucleon's spin is 
carried by the quarks~\cite{Bau83}. 
 
The integral Eq.~(\ref{SR8}) for the second spin structure functions
$G_2$ shows distinct differences in comparison with Eq.~(\ref{SR7}).
First, the helicity cross sections $\sigma_{1/2}$ and $\sigma_{3/2}$
appear with different sign. This has the consequence that in the sum
$I_{1+2} = I_1+I_2$ only the longitudinal-transverse contribution
$\sigma'_{LT}$ remains. Second, the latter contribution now appears as
$\frac{\nu}{Q}\sigma'_{LT}$, which is finite in the real-photon limit,
$Q^2\rightarrow 0$.  While the generalized GDH integral of
Eq.~(\ref{SR7}) ist not a sum rule, i.e. not related to another
observable except for the real photon point, the Burkhardt-Cottingham
(BC) sum rule predicts that $I_2(Q^2)$ can be expressed by the
magnetic $(G_M)$ and electric $(G_E)$ Sachs form factors at each
momentum transfer~\cite{Bur70},
 
\begin{equation} 
I_2(Q^2) = \frac{1}{4} G_M(Q^2) \frac{G_M(Q^2)-G_E(Q^2)}{1+Q^2/4m^2}\ . 
\label{SR9} 
\end{equation} 

According to Eq.~(\ref{SR9}) the integral $I_2$ approaches the value
$\kappa\mu/4$ for real photons $(Q^2=0)$ and drops with $Q^{-10}$ for
$Q^2\rightarrow\infty$. As a result the sum $I_{1+2}(0)$ should take
the value $\kappa(\mu-1)/4$, i.e. $\kappa^2/4$ and 0 for proton and
neutron respectively. However, there are strong indications that the
BC integrand gets large contributions at higher energies, which in
fact will affect its convergence. At least for the proton, however,
the ``sum rule'' seems to work quite well if we restrict ourselves to
the resonance region.
 
In the case of the proton, the GDH sum rule predicts $\Gamma_1<0$ for
small $Q^2$, while all experiments for $Q^2>1$ (GeV/c)$^2$ yield
positive values. Clearly, the value of the sum rule has to change
rapidly at low $Q^2$, with some zero-crossing at $Q^2_0<1$
(GeV/c)$^2$. This evolution of the sum rule was first parametrized by
Anselmino et al.~\cite{Ans89} in terms of vector meson dominance.
Burkert, Ioffe and others~\cite{Bur92} refined this model considerably
by treating the contributions of the resonances explicitly. Soffer and
Teryaev~\cite{Sof93} suggested that the rapid fluctuation of $I_1$
should be analyzed in conjunction with $I_2$, because $I_1+I_2$ is
known for both $Q^2=0$ and $Q^2 \rightarrow \infty$.  Though this sum
is related to the practically unknown longitudinal-transverse
interference cross section $\sigma'_{LT}$ and therefore not yet
determined directly, it can be extrapolated smoothly between the two
limiting values of $Q^2$. The rapid fluctuation of $I_1$ then follows
by subtraction of the BC value of $I_2$. We also refer the reader to a
recent evaluation of the $Q^2$-dependence of the GDH sum rule in a
constituent quark model~\cite{MaL98}, and to a discussion of the
constraints provided by chiral perturbation theory at low $Q^2$ and
twist-expansions at high $Q^2$ (see Ref.~\cite{JiO99}).

\begin{figure}[htbp] 
\centerline{\epsfxsize=10cm \epsfbox{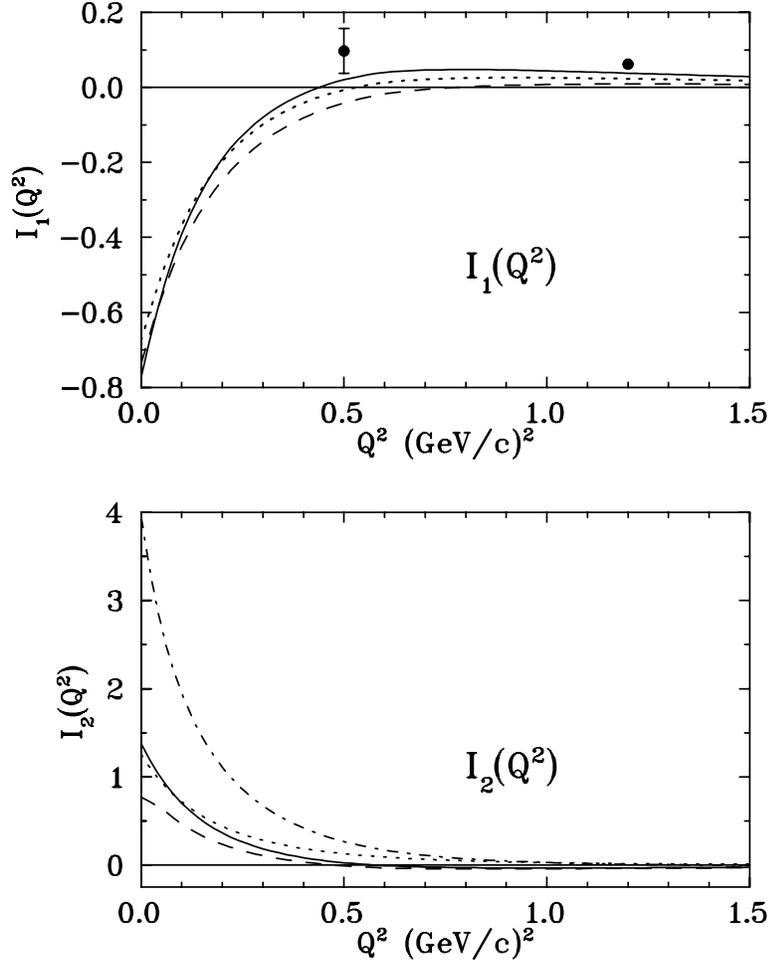}} 
\caption 
{The integrals $I_1$ and $I_2$ defined by Eqs.~(\ref{SR7})
  and~(\ref{SR8}) as functions of $Q^2$ in the resonance region,
  integrated up to $W_{max}=2$~GeV. Upper figure: Dashed, dotted and
  solid curves are calculations obtained with $1\pi, 1\pi+\eta$, and
  $1\pi+\eta+n\pi$ contributions, and data from
  Ref.~\protect\cite{Abe98}.  Lower figure: The full and dashed lines
  are our predictions~\protect\cite{Dre99b} with and without
  $\sigma'_{LT}$ [see Eq.~(\ref{SR8})], the dash-dotted line is
  obtained for the estimate $\sigma'_{LT}=\sqrt{\sigma_L\sigma_T}$,
  and the dotted line is the sum rule prediction of
  Ref.~\protect\cite{Bur70}.  All calculations for $I_2$ include
  $1\pi+\eta+n\pi$ contributions.}
\label{figSR15} 
\end{figure} 
 
In Fig.~\ref{figSR15} we give our predictions~\cite{Dre99b} for the
integrals $I_1(Q^2)$ and $I_2(Q^2)$ in the resonance region, i.e.
integrated up to $W_{max}=2\,GeV$. As can be seen, our model is able
to generate the dramatic change in the helicity structure quite well.
While this effect is basically due to the single-pion component
predicted by the UIM, the eta and multipion channels are quite
essential to shift the zero-crossing of $I_1$ from
$Q^2=0.75\,(GeV/c)^2$ to 0.52 $(GeV/c)^2$ and 0.45 $(GeV/c)^2$,
respectively.  This improves the agreement with the SLAC
data~\cite{Abe98}. However, some differences remain. Due to a lack of
data in the $\Delta$ region, the SLAC data are likely to underestimate
the $\Delta$ contribution and thus to overestimate the $I_1$ integral
or the corresponding first moment $\Gamma_1$. A few more data points
in the $\Delta$ region would be very useful in order to clarify the
situation, and we are looking forward to the results of the current
experiments at Jefferson Lab~\cite{Bur91}.
 
Concerning the integral $I_2$, our results are in good agreement with 
the predictions of the BC sum rule (see Fig.~\ref{figSR15}, lower part). 
The remaining differences are of the order of $10~\%$ and should be 
attributed to contributions beyond $W_{\rm max}=2$ GeV and the scarce 
experimental data for $\sigma'_{LT}$. 
 
We recall at this point that the results mentioned above refer to the
proton. Unfortunately, we find some serious problems for the neutron,
for which our model predicts both $I_1(0)$ and $I_2(0)$ larger
than expected from the sum rules. This has the consequence that our
prediction for $I_{1+2}(0)$ has a relatively large positive value
while it should vanish by sum rule arguments. The reason for this
striking discrepancy could well be due to the discussed problems with
``neutron targets''.  On the other hand it could also be an indication
of sizeable contributions at the higher energies, which could possibly
cancel for the proton but add in the case of the neutron. In this
context it is interesting to note that a recent parametrization of
deep inelastic scattering predicts sizeable high-energy contributions
with different signs for proton and neutron~\cite{Bia99}.
 
A more general argument is that the convergence of sum rules cannot be
given for granted, and thus the good agreement of our model with the
BC sum rule could be accidental and due to a particular model
prediction for the essentially unknown longitudinal-transverse
interference term. As can be seen from Fig.~\ref{figSR15}, the
contribution of $\sigma_{LT'}$ is quite substantial for $I_2$ even at
the real photon point due to the factor $\nu/Q$ in Eq.~(\ref{SR8}).
This contribution, however, is constrained by the positivity relation
$|\sigma'_{LT}|\le \sqrt{\sigma_L\sigma_T}$. The dash-dotted line
shows the integral for the upper limit of this inequality and a
similar effect would occur for the lower limit. The surprisingly large
upper limit can be understood in terms of multipoles. In a realistic
description of the integrated cross section $\sigma_{LT'}$, the large
$M_{1+}$ multipole can only interfer with the small $L_{1+}$
multipole. The upper and lower limits of the positivity relation
overestimate the structure function considerably due to an unphysical
``interference'' between $s$ and $p$ waves.

\section{SUMMARY}

The new generation of electron accelerators with high energy, 
intensity and duty-factor has made it possible to perform new classes 
of coincidence experiments involving polarization observables. These 
investigations have already provided new data with unprecedented 
precision, and they will continue to do so for the years to come. Some 
of the interesting topics and challenging questions are 
 
\begin{itemize} 
\item a full separation of the electric and magnetic form factors of 
  neutron and proton by double-polarization experiments, 
\item the search for strange quarks in the nucleon by parity-violating 
  electron scattering, 
\item new and more precise information on the scalar and vector 
  polarizabilities of the nucleon by a combined analysis of Compton 
  scattering and photoproduction as well as extensions to generalized 
  polarizabilities via virtual Compton scattering, 
\item the threshold amplitudes for the production of Goldstone bosons 
  and tests of chiral field theories, 
\item the quadrupole strength for $\Delta_{1232}$ excitation as a 
  measure of tensor correlations among the constituents, 
\item photo- and electroexcitation of the higher resonances, e.g. the 
  $N^{\ast}_{1440}$ (Why does the Roper occur at such a low excitation 
  energy? Where is its Coulomb monopole strength?), the 
  $N^{\ast}_{1535}$ (Is it really a resonance or a threshold effect of 
  $\eta$ production?), and the helicity structure of the main dipole 
  $(N^{\ast}_{1520})$ and quadrupole ($N^{\ast}_{1680})$ resonances 
  for both proton and neutron, 
\item investigations of individual decay channels including energy and 
  angular distributions in order to find out how much of the 
  excitation strength is actually due to resonances as opposed to 
  background and threshold effects, and more generally the question 
  how to extract the ``intrinsic'' quark structure from the 
  experimental data, which necessarily contain the hadronization in 
  terms of mesons, 
\item ongoing experiments to determine the helicity structure of 
  photo- and electroproduction in the resonance region by use of 
  double-polarization observables, which in turn are related to deep 
  inelastic scattering and the quark spin structure by means of sum 
  rules and related integrals over the excitation spectrum. 
\end{itemize} 
 
Our present understanding of nonperturbative QCD is still in a 
somewhat deplorable phenomenological state. The ongoing experimental 
activities will change that situation within short by providing new 
and detailed information on low-energy QCD in general and the 
nucleon's structure in particular. This rich phenomenology will 
without doubt challenge the theoretical approaches and, as is strongly 
to be hoped, eventually pave the way for a more quantitative 
understanding of nonperturbative QCD.

\newpage

{\bf TESTS AND PROBLEMS} 

\begin{enumerate} 
\item[I]  KINEMATICS
\item[1.] Prove that $s+t+u=m_1^2+m_2^2+m_3^2+m_4^2$, for the reaction 
  $p_1+p_2\rightarrow p_3+p_4$. 
\item[2.] Calculate the threshold energy in the lab frame for the 
  reactions 
\begin{itemize} 
\item[a)] 
          $p(\gamma, \gamma')p'$ 
\item[b)] $p(\gamma, \pi)p'$ 
\item[c)] replace the incident real photon by a virtual one with\\ 
  $m_{\gamma^{\ast}}^2=-Q^2<0$ 
\end{itemize} 
\item[3.] Which energy should an accelerator have to 
          electroproduce a $K^+$ at $Q^2=1$~(GeV/c)$^2$? How 
          much energy would you like to have before you schedule such 
          an experiment? 
\item[4.] In case of the reaction $p_1+p_2\rightarrow p_3+p_4$ 
          with 4 scalar particles, how many Lorentz scalars and 
          vectors can be constructed? How many are independent? 
\item[5.] Find the kinematical limits for Compton scattering 
          in the s-channel. Use $\nu=(s-u)/4M$ and $t$ as orthogonal 
          coordinates, and relate them in the $cm$ frame for forward 
          and backward scattering. What about the other parts of the 
          hyperbola of Fig.~3? 

\vspace{0.3cm} 
\item[II] FORM FACTORS 
 \item[6.] Evaluate the vector current $J_u=\bar{u}_{p'} 
  (F_1\gamma_{\mu}+\frac{i\sigma_{\mu\nu}q^{\nu}}{2m} F_2) u_p$ in the 
  Breit frame, $\vec{p}=-\frac{1}{2}\vec{q}$ and $\vec{p}\ 
  '=+\frac{1}{2}\vec{q}$, and identify the Sachs form factors $G_E$ 
  and $G_M$ of Eq.~(23). 
\item[7.] The electric Sachs form factor of the proton can be 
  approximated by the dipole form 
  $G_E^p=(1+(Q/.84$~GeV)$^2)^{-2}=G_D(Q^2)$.  Calculate the charge 
  distribution by a Fourier transform according to Eq.~(25). 
 \vspace{0.3cm} 
\item[III] STRANGENESS 
 \item[8.] Derive the structure of the Lorentz tensor 
  $W_{\mu\nu}=J_{\mu}J_{\nu}^{\ast}$, constructed from a general 
  vector current $J_{\mu}$ for an unpolarized fermion. Use independent 
  4-momenta $q_{\mu}=p_{2\mu}-p_{1\mu}$ and 
  $P_{\mu}=\frac{1}{2}(p_{1\mu}+p_{2\mu})$, and impose current 
  conservation. 
\item[9.] In the case of $\vec{e}+p\rightarrow e'+p'$, there also 
  appears an antisymmetric tensor 
  $\eta_{\mu\nu}^{({\cal{A}})}=\frac{i}{2m^2} 
  \epsilon_{\mu\nu\alpha\beta}q^{\alpha}P^{\beta}$ with a $\pm$ sign 
  in front depending on the helicity of the electron.  Why does this 
  term not contribute to the cross section derived from photon 
  exchange? What would be required to see this term? 
\item[10.] A simple model of the proton says that part of the time it 
  appears as a neutron surrounded by a $\pi^+$ cloud. If this system 
  has orbital momentum $l=l_z=1$, with which probability should it 
  occur in order to describe the anomalous magnetic moment $\kappa_p$ 
  of the proton? 
\item[11.] Estimate the contribution of $\Lambda^{\circ}K^+$ 
  configurations to the strangeness form factors of Eq.~(32) following 
  the procedure of No~10. Which sign does the strangeness magnetic 
  moment $\mu^s$ carry according to the model? What about the 
  strangeness radius $<r^2>_s$? 
 \vspace{0.3cm} 
\item[IV] COMPTON SCATTERING 
 \item[12.] Derive an upper limit for the electric polarizability of 
  proton and neutron in a two-body model as in No~10. Use the 
  definition of Eq.~(45) and note that the excitation spectrum lies at 
  $E_n-E_0>m_{\pi}$. Express the result in terms of the radius 
  $<r^2>_E^{n,p}$. How general is the result? 
\item[13.] A generic model of a polarizable system is the following
  (see Ref.~\cite{NatBo}): Two objects with masses $M_{1,2}$ and
  charges $Q_{1,2}$ are bound by a spring (oscillator frequency
  $\omega_0^2=C/\mu$, $C=$ Hooke's constant, $\mu=$ reduced mass). An
  electric $\vec{E}=\vec{E}_0e^{i\omega t}$ induces a dipole moment
  $\vec{D}=\alpha(\omega)\vec{E}_0$.
\begin{itemize} 
\item[a)] Determine $\alpha(\omega)$ and consider the cases of (I) 
  equal particles, (II) $M_2\rightarrow \infty$, $Q_2\rightarrow 0$. 
\item[b)] If $Q=Q_1+Q_2\neq 0$, the system will be accelerated even in 
  the limit $\omega\rightarrow 0$. Calculate 
  $\vec{D}^{\bullet\bullet}$ for the $cm$ coordinate. 
\item[c)] Calculate $\vec{D}^{\bullet\bullet}$ for the relative 
  coordinate. 
\item[d)] Classical antenna theory says that the cross section is 
 \[ 
  \frac{d\sigma}{d\Omega} = 
  |f(\omega)|^2 \sim \left 
    (\frac{|\vec{D}^{\bullet\bullet}|}{|\vec{E}|}\right )^2 \ . 
 \] 
 Compare this result with Eq.~(48) and discuss the scattering 
 amplitude $f$ and the cross section for $\omega=0$, $\omega\ll 
 \omega_0$, $\omega\approx \omega_0$ and $\omega\gg \omega_0$. Which 
 kind of scattering occurs for $Q=0$? 
\end{itemize} 
\item[14.] Estimate the polarizability for the following systems,
  approximated by 2-body configurations, using the result of No~13
  (see Ref.~\cite{NatBo}).
\begin{itemize} 
\item[a)] H atom = $p+e^-$ ,\ \ \ \  $\hbar\omega_0\sim 10$~eV 
\item[b)] deuteron = $p+n$ , \ \ \ $\hbar\omega_0\sim 4.5$~eV 
\item[c)] $^{208}Pb = 82p+126n$ , $\hbar\omega_0\sim 14$~eV 
\item[d)] $p=2u+1d$ , \ \ \ \ \ \ \ \ \ \  $\hbar\omega_0\sim 500$~eV 
\item[e)] $n = 1u+2d$ , \ \ \ \ \ \ \ \ \ \ $\hbar\omega_0\sim 500$~eV 
\end{itemize} 
\vspace{0.3cm} 
\item[V] PION PHOTOPRODUCTION 
 \item[15.] Threshold pion production is given by the s-wave multipoles 
  $E_{0+}$. Estimate these multipoles for the 4 physical channels 
  $\gamma+{\cal{N}}\rightarrow\pi+{\cal{N}}$ by evaluating the squares 
  of the electric dipole moments of the $\pi{\cal{N}}$ configuration. 
  Compare these results with Table~3. 
\item[16.]  Which multipoles connect the ${\cal{N}}$ with the 
  following resonances:\\ $P_{33}(1232),\ J^p=\frac{3}{2}^+$; 
  $P_{11}(1440),\ J^p=\frac{1}{2}^+;\\D_{13}(1520),\ 
  J^p=\frac{3}{2}^-;\ S_{11}(1535)$, $ J^p=\frac{1}{2}^-;\\
  F_{15}(1680),\ J^p=\frac{5}{2}^+$.\\See Eq.~(67) and the text 
  following that equation. 
\vspace{0.3cm} 
\item[VI] SUM RULES 
 \item[17.] The integrand of the GDH sum rule has the multipole 
  decomposition of Eq.~(76). Draw a figure of the GDH integrand for 
  the 4 physical channels as function of $\omega$, using the result of 
  No.~16 and the information given in Section 6. 
\item[18.] A possible generalization $I_{GDH}$ of the GDH integral is 
  obtained from Eq.~(83) by dropping the term in $\sigma_{LT}'$. 
  Discuss the sign of $I_{GDH}$ for $Q^2\Rightarrow 0$ and 
  $Q^2\Rightarrow\infty$. Give qualitative arguments why the sign 
  change takes place already at relatively small $Q^2$. 
\end{enumerate} 
 
 
\end{document}